\documentclass[submission, Phys]{SciPost}

\usepackage{physics}
\usepackage{amssymb}
\usepackage{amsmath}
\usepackage{enumitem}   
\usepackage{comment}
\usepackage{esint}

\hypersetup{
    colorlinks=true,
    linkcolor=blue,
    urlcolor=magenta,
    citecolor=red,
    }

\def\XXint#1#2#3{{\setbox0=\hbox{$#1{#2#3}{\int}$}
     \vcenter{\hbox{$#2#3$}}\kern-.5\wd0}}

\newcommand{\Qp}{Q_{\rm p}}
\newcommand{\Qap}{Q_{\rm ap}}

\newcommand{\sg}{s}

\newcommand{\vr}{\varrho}
\newcommand{\kbar}{\bar{k}}
\newcommand{\jj}{j_0}
\newcommand{\ww}{\Delta}

\begin{document}

\begin{center}{\Large \textbf{
Open quantum spin chains with non-reciprocity:\\
a theoretical approach based on\\
the time-dependent generalized Gibbs ensemble
}}\end{center}

    \begin{center}
Alice~Marché \textsuperscript{1*},
Hironobu~Yoshida \textsuperscript{2},
Alberto~Nardin \textsuperscript{1},
Hosho~Katsura \textsuperscript{2,3,4},
Leonardo~Mazza \textsuperscript{1,5}
\end{center}

\begin{center}
{\bf 1} Université Paris-Saclay, CNRS, LPTMS, 91405, Orsay, France
\\
{\bf 2} Department of Physics, Graduate School of Science, The University of Tokyo, 7-3-1 Hongo, Tokyo 113-0033, Japan
\\
{\bf 3} Institute for Physics of Intelligence, The University of Tokyo, 7-3-1 Hongo, Tokyo 113-0033, Japan
\\
{\bf 4} Trans-scale Quantum Science Institute, The University of Tokyo, 7-3-1, Hongo, Tokyo 113-0033, Japan
\\
{\bf 5} Institut Universitaire de France, 75005, Paris, France
\\
* alice.marche@universite-paris-saclay.fr

\end{center}

\begin{center}
\today
\end{center}

%For convenience during refereeing: line numbers
%\linenumbers

\section*{Abstract}
{\bf
We study an open quantum spin chain with non-reciprocal dissipation using a theoretical approach known as time-dependent generalized Gibbs ensemble.
In the regime of weak dissipation the system is fully characterized by its momentum distribution and we derive a closed set of coupled differential equations governing their time evolution. We check the accuracy of this theory by benchmarking the results against numerical simulations.
Using this framework we are able to compute both the magnetization density and current dynamics, identifying some relations between the two.
The problem of the anomalous power-law exponents identified in a previous work is discussed.
Our work constitutes a theoretical approach that is able to describe the physics of non-reciprocal open quantum spin chains beyond analyses based on non-interacting fermions.
}

\vspace{10pt}
\noindent\rule{\textwidth}{1pt}
\tableofcontents\thispagestyle{fancy}
\noindent\rule{\textwidth}{1pt}
\vspace{10pt}
%%%%%%%%%%%%%%%%%%%%%%%%%%%%%%%%%%%%%%%%%%%%%%%%%%%%%%%%%%%%%%%%%%%%%%%%%%%%%%%%%%%%%%%%%%%%%%%%%%%%%%%%%%%%%%%
%%%%%%%%%%%%%%%%%%%%%%%%%%%%%%%%%%%%%%%%%%%%%%%%%%%%%%%%%%%%%%%%%%%%%%%%%%%%%%%%%%%%%%%%%%%%%%%%%%%%%%%%%%%%%%%
\section{Introduction} \label{sec:intro}

In the classical world, out-of-equilibrium many-body systems often exhibit non-reciprocal interactions among their elementary constituents: the action-reaction symmetry is broken whenever one component can influence another without being affected in the same way.
Biological systems characterised by predator–prey interactions are a paradigmatic example, nowadays falling into the broad class of living active matter~\cite{Nagy2010,Tan2022,Dinelli2023}; recently, also synthetic systems have shown this behaviour~\cite{Uchida2010,Brandenbourger2019}.
Peculiar out-of-equilibrium phenomenology without a counterpart in equilibrium systems~\cite{Fruchart2021,Avni2025} has been identified, including spontaneous directed motion without external bias~\cite{Vicsek_2012}, odd elastic and transport responses~\cite{Fruchart_2023}, and unusual collective modes that violate standard fluctuation–dissipation relations~\cite{johnsrud_2025}.

Motivated by these results, the question of whether non-reciprocal quantum systems can be engineered has attracted significant attention.
Early studies of quantum non-reciprocity typically postulated a non-Hermitian Hamiltonian, which can arise from postprocessing a measured quantum system, or from a suitable integration of the environment~\cite{Ashida_2020}.
In this respect, the Hatano-Nelson model is a paradigmatic example, showcasing remarkable phenomena, such as the non-Hermitian skin effect, a strong spectral sensitivity on the boundary conditions, and non-reciprocal transport~\cite{Hatano1996,Hatano1997,Gong2018,Okuma2020}.

More recently, it has been proposed to study non-reciprocal interactions using a Lindblad master equation~\cite{Metelmann2015,Metelmann2017,Clerk2022}, which describes quantum systems weakly-coupled to a Markovian environment~\cite{Fazio2025}.
The main experimental proposal for realizing a non-reciprocal open quantum spin chain relies on two key aspects: (i)~the breaking of time-reversal symmetry, and (ii)~the presence of dissipation;
the interplay between these ingredients enables asymmetric couplings between quantum degrees of freedom~\cite{Metelmann2015,Metelmann2017,Clerk2022}.
In practice, one considers a  
nearest-neighbour Hamiltonian with synthetic gauge fields which break time-reversal invariance, while coupling neighbouring lattice sites to the same dissipative reservoir; the setup is then effectively described by a Lindblad master equation featuring two-site and spatially-directed jump operators.
The relation of these setups to the Hatano-Nelson non-Hermitian Hamiltonian, and the many similarities between the two, have been discussed~\cite{McDonald2022}.

This proposal has attracted a lot of interest and the model has been subsequently thoroughly investigated, focusing on the original bosonic formulation motivated by photonics, but also extending it to quantum spin chains and fermionic systems.
The appearance of persistent currents induced solely by reservoir engineering has been clearly demonstrated~\cite{Keck2018}.
In the presence of hard-wall boundary conditions, the setup displays eigenstates that are exponentially localized at the edges, a manifestation of the non-Hermitian skin effect~\cite{Song_2019}.
This phenomenology has also been employed to discuss directional amplification in non-reciprocal open quantum systems~\cite{Porras_2019, Wanjura_2020, Brunelli_2023}.
The existence of repulsively-bound doublon quasiparticles in a bosonic setting motivated by quantum-optics applications, which are bound together by interactions, and their non-reciprocal dynamics, has been shown in Ref.~\cite{Brighi2024}.

Other works have specifically focused on the emergence of universal critical behaviors in the late-time algebraic decay $\sim t^{-\chi}$ of an observable associated with the closure of the Lindbladian spectral gap in the thermodynamic limit~\cite{Begg2024, Belyansky2025, Soares2025}.
Different power law exponents $\chi$ have been found depending on whether one considers (i) a fermionic system, focusing on the fermionic number, or (ii) a spin system, discussing the magnetization.
In the case (i), a clean power law exponent $\chi=1/2$ is found both numerically and analytically. In contrast, for the case (ii), analytical derivations are much more challenging, and the study of the system relies on numerical simulations. By fitting numerical data, one observes that the exponent $\chi$ depends on both the initial state and the strength of non-reciprocity; for instance $\chi = 0.58$ or $\chi =0.515$ have been reported~\cite{Begg2024}.
This non-trivial behavior remains poorly understood and calls for the development of new theoretical frameworks.

In this article, we contribute to the studies on open and non-reciprocal quantum spin chains by proposing a theoretical approach based on the time-dependent generalized Gibbs ensemble (t-GGE)~\cite{Lange2018}.
This framework applies to systems which are integrable in the absence of dissipation and are expected to thermalize to a generalized Gibbs ensemble (GGE) which maximizes the entropy while satisfying the extensive number of conservation laws of the Hamiltonian~\cite{VidmarRigol2016, Essler_2016}.
When a weak dissipation is introduced, thereby breaking the conservation laws, the t-GGE approach postulates that the system should be described by a time-dependent GGE.
This formalism has been used to describe various types of Lindbladian evolutions, such as lossy one-dimensional gases~\cite{Bouchoule2020, Rossini2021, Rosso2022, Riggio2024, Lumia2025}, reaction--diffusion dynamics in quantum gases~\cite{Perfetto_2023, Lehr2025}, digital quantum processors~\cite{Ulcakar2025}, trapped ions~\cite{Reiter_2021}, fermionic chains with bath-induced currents~\cite{Yamamoto2020}.
What makes t-GGE a well-established tool is that its predictions have been successfully validated through tensor-network simulations.

We show that the momentum distribution is a crucial theoretical object for understanding the dissipative behavior of a open non-reciprocal quantum spin chain.
For the XX model considered in this work, the occupation numbers of quasiparticles that carry energy and magnetization correspond to the momentum distribution of the fermions obtained after the Jordan--Wigner transform and define a complete set of extensive conserved quantities. This construction can be generalized to interacting systems by replacing the momentum distribution with a rapidity distribution.

When the coherent part of the dynamics is governed by the XX Hamiltonian, we derive a closed set of differential equations whose number scales linearly with the system size, making them significantly cheaper to solve than tensor-network simulations. From the momentum distribution, several physical observables can be directly computed, including the magnetization density and magnetization current, allowing us to formulate quantitative predictions and to provide a clear physical interpretation of the dynamics. Comparison with numerical simulations shows that this approach remains quantitatively accurate even in regimes where dissipation is not perturbative. We also formulate an analytical theory that captures the intermediate- and long-time evolution of the magnetization density and explains deviations from a purely algebraic decay. In summary, we demonstrate that the physics of a class of open, non-reciprocal quantum spin chains can be accessed by focusing solely on their momentum distribution.

This work is organized as follows.
In Sec.~\ref{Sec.Model} we introduce the model of the open and non-reciprocal quantum spin chain that we will consider;
we briefly review some known results and compare them with a simple theoretical analysis based on non-interacting fermions, which has several deficiencies.
Our theoretical approach based on the t-GGE is presented in Sec.~\ref{sec:GGE}, where we derive a closed set of differential equations governing the time evolution of the momentum distribution.

By solving numerically those equations, in Sec.~\ref{Sec:Magnetization}  we study the temporal decay of the magnetization of the chain, whereas the onset of a non-reciprocal current is discussed in Sec.~\ref{Sec:Current}.
In Sec.~\ref{Sec:AnalyticalCharacterization}, we characterize the late-time decay of the magnetization by deriving a closed dynamical equation for this quantity.
Our conclusions are presented in Sec.~\ref{Sec:Conclusions}.
Eight appendices conclude the work.
Throughout the article we set $\hbar=1$.

%%%%%%%%%%%%%%%%%%%%%%%%%%%%%%%%%%%%%%%%%%%%%%%%%%%%%%%%%%%%%%%%%%%%%%%%%%%%%%%%%%%%%%%%%%%%%%%%%%%%%%%%%%%%%%%
%%%%%%%%%%%%%%%%%%%%%%%%%%%%%%%%%%%%%%%%%%%%%%%%%%%%%%%%%%%%%%%%%%%%%%%%%%%%%%%%%%%%%%%%%%%%%%%%%%%%%%%%%%%%%%%
\section{Model and known results}
\label{Sec.Model}
%%%%%%%%%%%%%%%%%%%%%%%%%%%%%%%%%%%%%%%%%%%%%%%%%%%%%%%%%%%%%%%%%%%%%%%%%%%%%%%%%%%%%%%%%%%%%%%%%%%%%%%%%%%%%%%
%%%%%%%%%%%%%%%%%%%%%%%%%%%%%%%%%%%%%%%%%%%%%%%%%%%%%%%%%%%%%%%%%%%%%%%%%%%%%%%%%%%%%%%%%%%%%%%%%%%%%%%%%%%%%%%

%%%%%%%%%%%%%%----------------------------------------------------------------------------------%%%%%%%%%%%%%%
\subsection{The model} \label{sec:model}
%%%%%%%%%%%%%%----------------------------------------------------------------------------------%%%%%%%%%%%%%%

We consider a one-dimensional spin chain whose time evolution is governed by the XX~Hamiltonian
\begin{subequations}
\begin{equation}
H = -\frac{J}{2} \sum_j \left( S_{j+1}^+ S_{j}^- + S_{j}^+ S_{j+1}^- \right), \label{eq:XXhamiltonian}
\end{equation}
where $J$ is the exchange interaction, while $S_j^+ = (\sigma^x_j+i\sigma^y_j)/2$ and $S_j^-= (\sigma^x_j-i\sigma^y_j)/2$ denote the raising and lowering spin operators at the site $j$, respectively;
$\sigma^x$ and $\sigma^y$ are Pauli matrices.
We assume that this system is subject to two-site correlated dissipation as introduced in Ref.~\cite{Begg2024}: the time evolution of the system's density matrix $\rho$ is modeled by the Lindblad master equation
\begin{equation}
    \frac{d \rho}{ dt} = - i \left[ H , \rho\right] + \kappa \sum_{j} \left( L_j \rho L_j^\dag - \frac{1}{2} \{ L_j^\dag L_j , \rho \} \right) \label{eq:ME},
\end{equation}
where the jump operators reads
\begin{equation}
    L_j = S^-_j + e^{i \phi} S^-_{j+1}. \label{eq:jumps}
\end{equation}
Here, $\kappa >0$ is the loss rate and $\phi \in (-\pi,\pi]$. The jump operators~\eqref{eq:jumps} not only describe the decay of magnetization along the $z$-axis
but also account for the presence of non-reciprocal interactions, leading to a transient magnetization current
whenever $\phi \neq 0 \pmod{\pi}$~\cite{Keck2018, Begg2024}. The strength of the non-reciprocity is controlled via both the loss rate $\kappa$ and the phase~$\phi$; for fixed~$\kappa$, the non-reciprocity is maximal when $\phi = \pm \pi/2$ whereas the system is reciprocal when $\phi = 0,\pi$. \label{eq:model}
\end{subequations}
We will consider translationally-invariant initial states of the form
 \begin{equation}
  \ket{\Psi_0} = \bigotimes_{j} \left( \cos  \theta  \ket{\uparrow}_j + \sin \theta  \ket{\downarrow}_j \right), \quad  \text{ with } \theta \in [0, \frac{\pi}{2}). \label{eq:Psi0}
 \end{equation}
Unless otherwise specified, we consider an infinite number of lattice sites and the boundary conditions do not matter.
We denote the expectation value of an observable $O$ as $\langle O \rangle (t) := \Tr[O \rho(t)]$;
in the following, the explicit time dependence $(t)$ is often suppressed for brevity.
%%%%%%%%%%%%%%----------------------------------------------------------------------------------%%%%%%%%%%%%%%
\subsection{Known results on the algebraic decay of the magnetization} \label{sec:resultsBH}
%%%%%%%%%%%%%%----------------------------------------------------------------------------------%%%%%%%%%%%%%%
%%%%%%%------FIG---------%%%%%%%
\begin{figure}[t]
\centering
 \includegraphics[width=0.9\columnwidth]{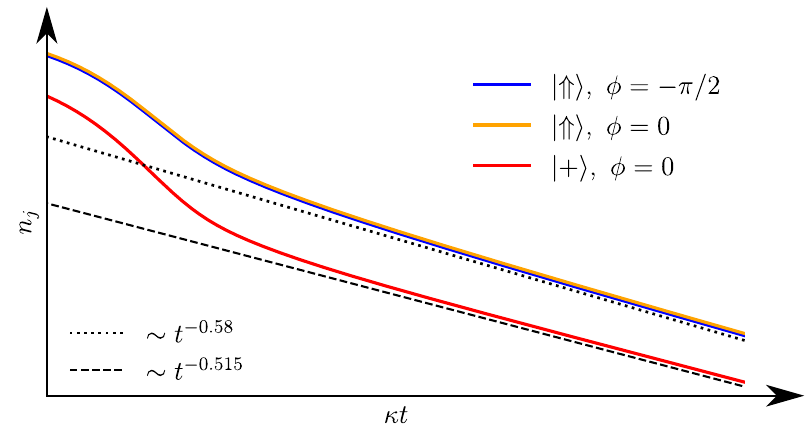}
 \caption{Sketch of the time evolution of the local magnetization $n_j$, defined in Eq.~\eqref{Eq:n:Spin}, for the two initial states $\ket{\Uparrow}$ and $\ket{+}$, in both the reciprocal ($\phi = 0$) and maximally non-reciprocal ($\phi = -\pi/2$) cases. The colored curves represent the results found in Ref.~\cite{Begg2024} via tensor-network (TN) simulations. The dotted and dashed blacks lines correspond to the fitted algebraic decays at late times.}
\label{fig:n_sketch}
\end{figure}
%%%%%%%-----------------%%%%%%%

The spin model, described by the Eqs.~\eqref{eq:model}, has been analyzed in Ref.~\cite{Begg2024} with tensor-network techniques.
The authors computed the late-time decay of the local magnetization~$S^z_j$ via the proxy projector onto the $\ket{\uparrow}$ state,
\begin{equation}
    n_j := S_j^+ S_j^- .
    \label{Eq:n:Spin}
\end{equation}
For open boundary conditions, and at late times, they observed an algebraic decay $\langle n_j \rangle \sim t^{-\chi}$ that depends on the initial state. By fitting the numerical data, they found $\chi = {0.58}$ when initially all spins are up $\ket{\Uparrow} = \bigotimes_j \ket{\uparrow}_j$, for both the reciprocal ($\phi=0$) and maximally non-reciprocal ($\phi=-\pi/2$) cases. On the other hand, they fitted $\chi = {0.515}$, for $\phi=0$, when initially all spins are aligned along the $x$-direction $\ket{+} = \bigotimes_j \left( \frac{\ket{\uparrow}_j +  \ket{\downarrow}_j}{\sqrt{2}} \right)$; their results are sketched in Fig.~\ref{fig:n_sketch}.

In general, the algebraic decays of observables are a consequence of the closure of the Lindbladian gap and the presence of a large number of slow modes in the thermodynamic limit~\cite{Medvedyeva_2014}: the system is then called critical. This phenomenon is widely observed in the literature~\cite{Cai2013, Poletti_2012, Poletti_2013, Marche2024, Soares2025, Fazio2025}. By empirical observation, the power law exponents are typically rational numbers of the form~$-p/q$ where both $p$ and $q$ are small (for instance $-1$ or $-1/2$). Therefore, these power law exponents $-0.58$ and $-0.515$ are rather unexpected.

For completeness, we mention that Ref.~\cite{Begg2024} is not the only system in which anomalous power-law exponents have been observed numerically (although this article will focus on this particular case). Similar behavior has also been found, for instance, in the decay of particle density in gases of hard-core bosons with three-body losses~\cite{Riggio2024}, as well as in the coherence decay of XXZ spin chains with on-site dephasing~\cite{Cai2013}.

\subsection{Free fermions}

%%%---FIG----%%%%
\begin{figure}[t]
 \includegraphics[width=\columnwidth]{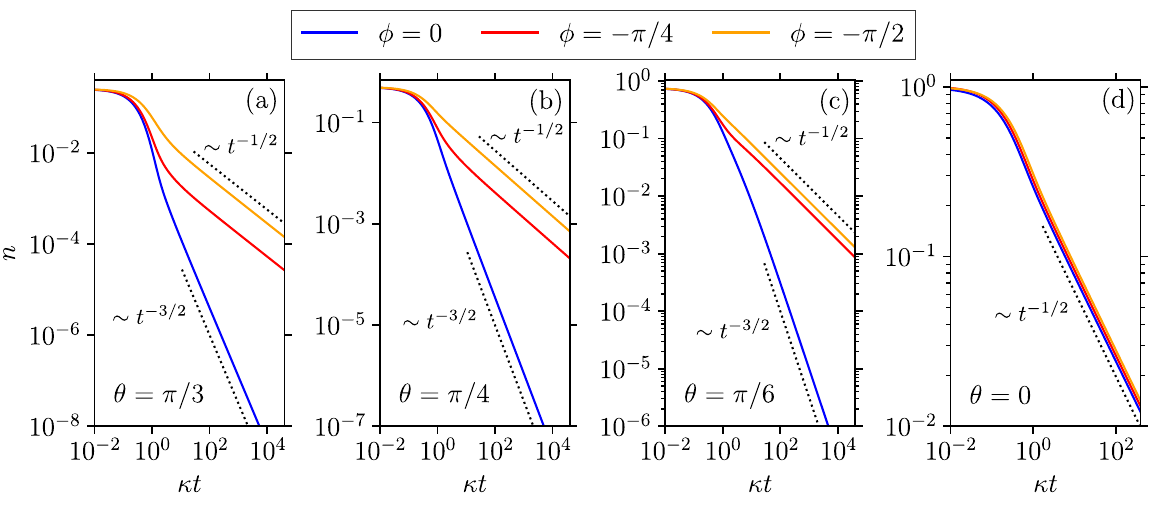}
 \caption{Time evolution of the local fermion density $n$ for the model~\eqref{eq:ff} for several non-reciprocity angles $\phi=0,-\pi/4,-\pi/2$.
 Different panels correspond to different initial states, parametrized by $\theta$ according to the expression in Eq.~\eqref{Eq:Initial:State}:
 panel (a), $\theta=\pi/3$;
 panel (b), $\theta=\pi/4$ corresponding to $\ket{\Psi_0} = \ket{+}$;
 panel (c), $\theta=\pi/6$;
 panel (d), $\theta=0$ corresponding to $\ket{\Psi_0} = \ket{\Uparrow}$.
 The colored continuous curves are obtained by numerically evaluating the integral~\eqref{eq:n_ff}. The dotted curves are guides to the eye for the late-time power-law
 scaling. In panel (d), all the curves overlap perfectly; the curves for $\phi=0,-\pi/4$ have been slightly shifted to improve visibility.
 }
\label{fig:ff}
\end{figure}
%%%---------%%%%
As a first step towards analyzing the spin model introduced above, we follow Refs.~\cite{Begg2024, Soares2025} and consider an exactly solvable free fermionic model whose Hamiltonian and jump operators are
\begin{equation}
 H = - \frac{J}{2} \sum_{j} \left( c_{j+1}^\dag c_j + c_j^\dag c_{j+1} \right), \qquad L_j = c_j + e^{i \phi} c_{j+1}. \label{eq:ff}
\end{equation}
Here, $c_j^\dag$ and $c_j$ denote the creation and annihilation operators of a fermion at site $j$. We consider this model because, as will be shown in the upcoming section, it is closely related to the one described by Eqs.~\eqref{eq:model}, even though not equivalent to it. To investigate the model~\eqref{eq:ff}, we introduce the Fourier modes
\begin{equation}
\forall k \in [0,2\pi), \ c^\dag(k) = \frac{1}{\sqrt{2 \pi}} \sum_{j = - \infty}^{+ \infty} e^{i k j} c_j^\dag, \qquad  c_j^\dag = \frac{1}{\sqrt{2 \pi}} \int_{0}^{2 \pi} e^{-ikj} c^\dag(k) dk,
\end{equation}
and the momentum distribution $\varrho(k)$ defined via
\begin{equation}
 \langle c^\dag(k) c(q) \rangle = \delta(k-q) \varrho(k). \label{eq:defrho}
\end{equation}
Note that $\langle c^\dag(k) c(q) \rangle = 0$ when $k\neq q$ is a direct consequence of the translational invariance of the system (see App.~\ref{ap:TI} for a proof). The initial state~\eqref{eq:Psi0} can be written in the fermionic basis using the Jordan-Wigner transform
\begin{equation}
 c_j =  (-1)^{N_{(-\infty,j-1)}} S_j^-, \qquad c_j^\dag = (-1)^{N_{(-\infty,j-1)}} S_j^+, \qquad \text{ with } \quad N_{(-\infty,j)} = \sum_{l = -\infty}^{j} n_l. \label{eq:JW}
\end{equation}
We then have
\begin{equation}
 \ket{\Psi_0} = \ldots \left( \cos \theta c_1^\dag + \sin \theta \right) \left( \cos \theta c_2^\dag + \sin \theta \right) \left( \cos \theta c_3^\dag + \sin \theta \right) \ldots \ket{\rm vac},
 \label{Eq:Initial:State}
\end{equation}
where $\ket{\rm vac} = \bigotimes_j \ket{\downarrow}_j$ is the vacuum state. The 
momentum distribution of $\ket{\Psi_0}$ is
\begin{equation}
    \varrho_0(k) = \begin{cases}   \dfrac{2 \cos^4(\theta) \left( 1 + \cos(k) \right) }{1+ 2 \cos(k)\cos(2 \theta) + \cos^2(2 \theta)}, & \text{ if } \theta \in (0,\frac{\pi}{2}), \\ \\   1, & \text{ if } \theta = 0. \end{cases} \label{eq:IC}
\end{equation}
 (see App.~\ref{ap:rho0} for details). Using the free fermionic nature of the model, we can show that~\cite{Begg2024,Soares2025}
\begin{equation}
 \frac{d}{dt} \varrho(k) = - 2 \kappa \left( 1 + \cos(\phi+k) \right) \varrho(k). \label{eq:rhok_ff}
\end{equation}
Hence, the time-dependent translationally-invariant density of fermions is 
\begin{equation}
 n(t) := \langle n_j \rangle(t) =  \langle c_j^\dag c_j \rangle(t) = \int_0^{2 \pi} \frac{dk}{2 \pi} \varrho(k,t) = \int_0^{2 \pi} \frac{dk}{2 \pi} \varrho_0(k) e^{- 2 \kappa \left( 1 + \cos(\phi+k) \right)t}. \label{eq:n_ff}
\end{equation}
From the Eqs.~\eqref{eq:IC} and~\eqref{eq:n_ff}, we can easily compute numerically the time evolution of $n$, as shown in Fig.~\ref{fig:ff}.
At late time, we observe power-law
behaviors 
\begin{equation}
  n \sim t^{-\chi} \quad \text{with} \quad  \chi = \begin{cases}  3/2, & \text{ if }  \phi=0 \text{ and } \theta \in (0,\frac{\pi}{2}), \\ 1/2, & \text{otherwise}. \end{cases} \label{eq:chi_ff}
\end{equation}
This can be shown analytically; see App.~\ref{ap:freefermions} for a proof.
Thus, already at the level of free fermions we find that different power-law exponents~$\chi$ emerge, depending both on the strength of the non-reciprocity and on the initial state. In particular, the configurations $(\phi = 0, \theta = 0)$ and $(\phi = -\pi/2, \theta = 0)$ yield the same value of $\chi$, whereas the case $(\phi = 0, \theta = \pi/4)$ does not, mirroring the behavior reported previously for the analogous spin system in Ref.~\cite{Begg2024}.
Once more, we observe that in this situation where an analytical calculation could be performed, the value of $\chi$ takes a simple rational form.

%%%%%%%%%%%%%%%%%%%%%%%%%%%%%%%%%%%%%%%%%%%%%%%%%%%%%%%%%%%%%%%%%%%%%%%%%%%%%%%%%%%%%%%%%%%%%%%%%%%%%%%%%%%%%%%
%%%%%%%%%%%%%%%%%%%%%%%%%%%%%%%%%%%%%%%%%%%%%%%%%%%%%%%%%%%%%%%%%%%%%%%%%%%%%%%%%%%%%%%%%%%%%%%%%%%%%%%%%%%%%%%
\section{Time-dependent generalized Gibbs ensemble (t-GGE)} \label{sec:GGE}
%%%%%%%%%%%%%%%%%%%%%%%%%%%%%%%%%%%%%%%%%%%%%%%%%%%%%%%%%%%%%%%%%%%%%%%%%%%%%%%%%%%%%%%%%%%%%%%%%%%%%%%%%%%%%%%
%%%%%%%%%%%%%%%%%%%%%%%%%%%%%%%%%%%%%%%%%%%%%%%%%%%%%%%%%%%%%%%%%%%%%%%%%%%%%%%%%%%%%%%%%%%%%%%%%%%%%%%%%%%%%%%
 In this section, we present our theoretical framework for the study of the spin model described by Eqs.~\eqref{eq:model}.
 Crucially, we assume that the system is weakly dissipative \textit{i.e.}~$\kappa \ll J$;
 as a consequence, the coherent unitary dynamics occurs on a much shorter time scale than that of the loss processes. In this context, it is customary to assume that the system relaxes locally to a time-dependent generalized Gibbs ensemble (t-GGE)~\cite{Lange2018,Bouchoule2020,Rossini2021, Riggio2024,Lehr2025,Ulcakar2025,Lumia2025}.

%%%%%%%%%%%%%%----------------------------------------------------------------------------------%%%%%%%%%%%%%%
\subsection{t-GEE for the spin model~\eqref{eq:model} on an infinite lattice}
%%%%%%%%%%%%%%----------------------------------------------------------------------------------%%%%%%%%%%%%%%

First, we map the spin operators $S_j^+$, $S_j^-$ to the fermionic operators $c_j^\dag$, $c_j$ via the Jordan-Wigner transform~\eqref{eq:JW}.
Under this transformation, the Hamiltonian and the jump operators can then be rewritten as
\begin{equation}
  H = -\frac{J}{2} \sum_{j = -\infty}^{+ \infty} \left( c_{j}^\dag c_{j+1} + c_{j+1}^\dag c_j \right), \quad L_j  = (-1)^{N_{(-\infty,j-1)}} \left( c_j + e^{i \phi} \left( -1 \right)^{n_j} c_{j+1} \right) \label{eq:jumps_fermions}.
\end{equation}
We note that our spin model is different from the free fermions~\eqref{eq:ff} because of the strings $(-1)^{N_{(-\infty,j-1)}}$ and $(-1)^{n_j}$ appearing in $L_j$.
This is the essence of the difference between the two approaches; the importance of Jordan-Wigner strings in open quantum system dynamics has been recently highlighted in Ref.~\cite{Pocklington_2025}.

The Hamiltonian is diagonal in Fourier space:
\begin{equation}
 H = \int_0^{2 \pi} \epsilon_k c^\dag(k) c(k) dk, \quad \text{with} \quad \epsilon_k = - J \cos k. \label{eq:HFourier}
\end{equation}
Thus, we can identify the 
mutually commuting and extensive operators which also commute with $H$, also known as the extensive charges. Denoting them as $\{ Q_m \}$, we have $[Q_m,H] = [Q_m,Q_n] = 0$. We recall that, by definition, if $Q_m$ is extensive then it can be written as $Q_m = \sum_j q_{j,m}$ where $q_{j,m}$ is its density operator and acts on a finite number of sites around $j$ (\textit{i.e.}~has a finite support).
From the expression of $H$ in Eq.~\eqref{eq:HFourier}, we see that the charges can be expressed as
\begin{subequations}
\begin{align}
    Q_{2 p} =& \int_0^{2 \pi} \sin(p k) c^\dag(k) c(k) dk,
    \\
    Q_{2 p+1} =& \int_0^{2 \pi} \cos(p k) c^\dag(k) c(k) dk;
    \qquad  \forall p \in \{0,1,2,\ldots\}. \label{eq:CC}
\end{align}
\end{subequations}
In the absence of losses, \textit{i.e.}~$\kappa=0$, the system would relax to a generalized Gibbs ensemble (GGE) characterized by $\{ Q_{m} \}_{m>0}$. However, in the case of weak adiabatic losses, the integrability is weakly broken, meaning that the GGE is not exactly stationary. Therefore, we assume that the system's density matrix is locally indistinguishable from
\begin{equation}
     \rho_{\rm GGE}(t) = \frac{e^{-\sum_{m} \beta_{m}(t) Q_{m}}}{\Tr[e^{-\sum_m \beta_{m}(t) Q_{m}}]}, \label{eq:t-GGE}
\end{equation}
 where the Lagrange multipliers $\beta_{m}(t)$ are slowly varying in time.
 The state~\eqref{eq:t-GGE} is called a t-GGE and can also be expressed as
\begin{equation}
    \rho_{\rm GGE}(t) =  \frac{1}{\mathcal{Z}(t)} e^{-\int  \lambda_k(t) c^\dag(k) c(k) dk}, \label{eq:GGEstate}
\end{equation}
with
$\lambda_k(t) = \sum_p \beta_{2p+1} (t) \cos(p k)+   \beta_{2p} (t) \sin(p k) $ and  $\mathcal{Z}(t)$ is the appropriate normalization constant. Remarkably, the state~\eqref{eq:GGEstate} is a Gaussian fermionic state; this property will be used in order to apply the Wick theorem.

%%%%%%%%%%%%%%----------------------------------------------------------------------------------%%%%%%%%%%%%%%
\subsection{Equation of motion for the momentum distribution} \label{sec:rhok}
%%%%%%%%%%%%%%----------------------------------------------------------------------------------%%%%%%%%%%%%%%
We aim to derive a closed set of dynamical equations for the momentum distribution $\varrho(k)$ defined in Eq.~\eqref{eq:defrho}.
For this, we will follow similar steps as in Refs.~\cite{Bouchoule2020,Riggio2024}.
From the Lindblad master equation~\eqref{eq:ME}, we obtain
 \begin{equation}
\frac{d}{dt} \langle c^\dag(k) c(k) \rangle = - \kappa \sum_{j=-\infty}^{\infty} \left( \text{Re} \left[ \langle L_j^\dag L_j c^\dag(k) c(k) \rangle \right] - \langle L_j^\dag  c^\dag(k) c(k) L_j  \rangle \right). \label{eq:evolcc}
 \end{equation}
Using the Eqs.~\eqref{eq:defrho} and \eqref{eq:evolcc}, and the translational invariance, we find
\begin{equation}
- \frac{1}{ 2 \pi \kappa}\frac{d \varrho(k)}{dt} = \text{Re} \left[  \langle L_{\jj}^\dag L_{\jj} c^\dag(k) c(k) \rangle \right] - \langle L_{\jj}^\dag  c^\dag(k) c(k) L_{\jj} \rangle. \label{eq:evolcc2}
\end{equation}
Here, since the system is translationally invariant, $j_0$ can be any site index.

By assuming that the density matrix of the system  is locally indistinguishable from the t-GGE~\eqref{eq:GGEstate}, we can show
\begin{equation}
 \begin{split}
   - \frac{1}{2 \kappa} \frac{d \varrho(k)}{dt} =  \varrho(k) & \left( 1- \varrho(k) \right) \left( 1 + \cos(k + \phi) \right) + 2 \left(  \fint_{0}^{2 \pi} \frac{dq}{2 \pi} \varrho(q) \frac{\cos \left( \frac{q+\phi}{2}\right)}{\sin \left( \frac{k-q}{2}\right)}\right)^2 \\ &  +  \int_{0}^{2 \pi} \frac{dq}{2 \pi} \varrho(q)  \left( 1+\cos(q+\phi) \right) \fint_{0}^{2 \pi}  \frac{dp}{2 \pi} \frac{\varrho(k)- \varrho(p)}{\sin^2 \left( \frac{k-p}{2}\right)} .  \label{eq:GGEeq}
\end{split}
\end{equation}
The notation $\fint$ denotes the Cauchy principal value of the integral. Eq.~\eqref{eq:GGEeq} is the main result of this article; its complete derivation is provided in App.~\ref{ap:Ap1}.

In order to solve numerically the latter equation, we use the following alternative form, involving circular Hilbert transforms,
\begin{equation}
\begin{split}
    - \frac{1}{ 2 \kappa} \frac{d \varrho(k)}{dt}  & =   \left( \varrho(k) +  \varrho_c(k) \right) \left( 1- \varrho(k) \right) +  n  \left(  n + 2 \mathcal{H}[\varrho]^\prime(k)+    \mathcal{H}[\varrho_s](k)   +   \int_0^{2 \pi} \frac{dq}{2 \pi} \varrho_c(q) \right)
   \\ & +   \mathcal{H}[\varrho](k) \times \left( \mathcal{H}[\varrho](k) + \mathcal{H}[\varrho_c](k) -  \int_0^{2 \pi}  \frac{dp}{2 \pi} \varrho_s(p)\right) + 2 \mathcal{H}[\varrho]^\prime(k)  \int_0^{2 \pi} \frac{dq}{2 \pi} \varrho_c(q). \label{eq:GGEeqNum}
\end{split}
\end{equation}
Here,
\begin{equation}
 n = \int_0^{2 \pi} \frac{dk}{2 \pi} \varrho(k)
 \label{Eq:Def:Fermi:Dens}
\end{equation}
is the fermion density, which is related to the magnetization density via Eq.~\eqref{Eq:n:Spin}; we have used the short-hand notations $\varrho_c(k) = \varrho(k) \cos \left( k+ \phi  \right)$, $\varrho_s(k) = \varrho(k) \sin \left( k+ \phi  \right)$. In addition, we recall the definition of the circular Hilbert transform and the expression of its derivative~\cite{King2009,Pandey1997}:
\begin{equation}
 \mathcal{H}[f](k) = \fint_0^{2 \pi} \frac{dp}{2 \pi} \cot \left(\frac{k-p}{2} \right) f(p), \qquad  \mathcal{H}[f]'(k)  = \frac{1}{2} \fint_0^{2 \pi} \frac{dp}{2 \pi} \frac{f(k) - f(p)}{\sin^2 \left( \frac{k-p}{2}\right)}.
\end{equation}
The procedure used to solve numerically Eq.~\eqref{eq:GGEeqNum} is detailed in App.~\ref{ap:Ap2}.

\subsection{The time-evolving momentum distribution}
\label{SubSec:T:rhok}

%%%%%%%%%%%%%%%%%%%%%%%%%%%%%%%%
%%%%%%%------FIG---------%%%%%%%
\begin{figure}[t]
 \includegraphics[width=\columnwidth]{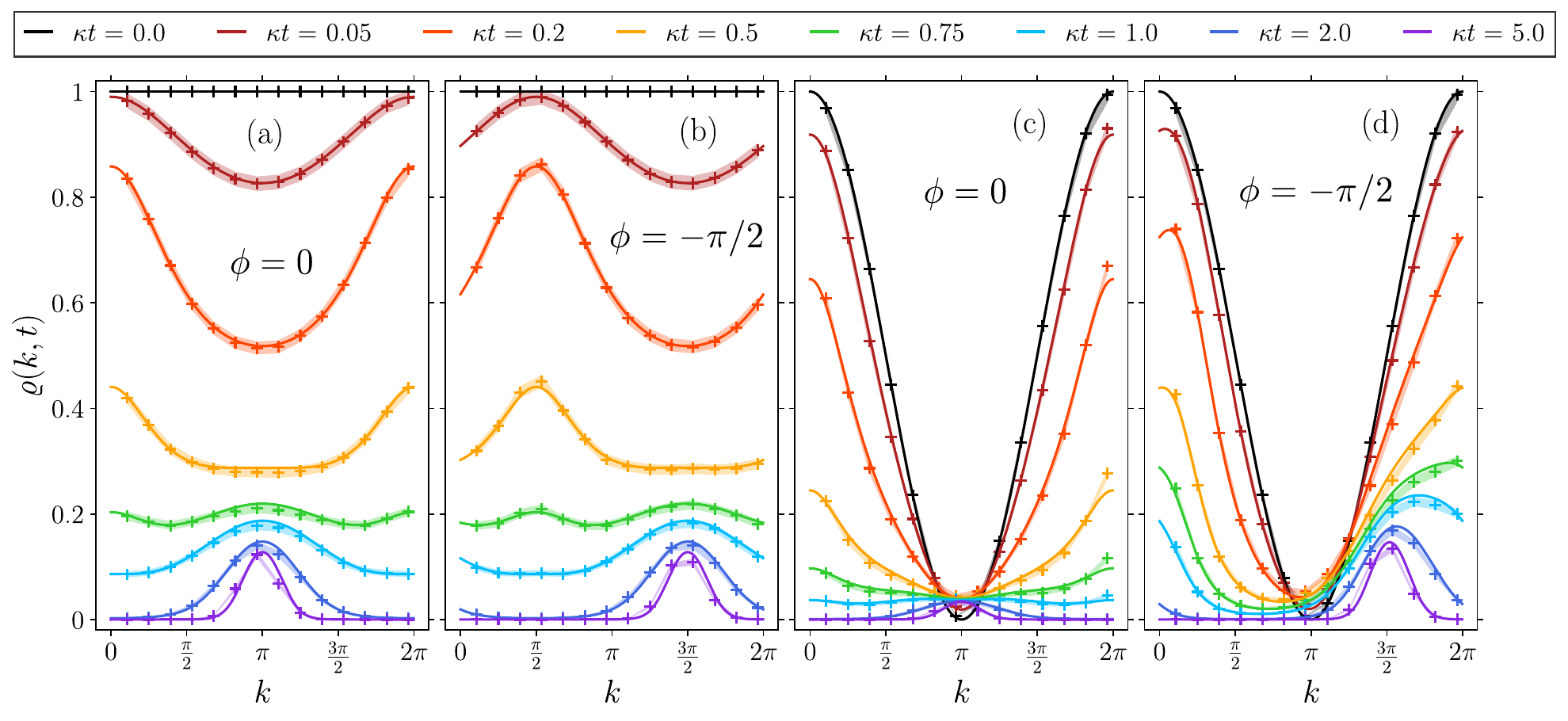}
 \caption{Time evolution of the momentum distribution $\varrho(k,t)$.
 Different panels refer to different initial states and angles $\phi$:
 (a) $\ket{\Uparrow}$ and $\phi=0$;
 (b) $\ket{\Uparrow}$  and $\phi=-\pi/2$;
 (c) $\ket{+}$  and $\phi=0$;
 (d) $\ket{+}$ and $\phi=-\pi/2$.
 For all panels, the continuous curves are obtained by solving numerically the Eq.~\eqref{eq:GGEeqNum}.
 On the other hand, the crosses correspond to a finite lattice of $L=14$ sites where computations are performed via a quantum trajectory stochastic algorithm, averaging over 5000 trajectories, with the parameters $J=1$ and $\kappa =0.02$, and periodic boundary conditions.}
\label{fig:rhok}
\end{figure}
%%%%%%%%%%%%%%%%%%%%%%%%%%%%%%%%

In Fig.~\ref{fig:rhok}, we show the distribution $\varrho(k)$ at various times, for two different initial states, $\ket{\Uparrow}$ and $\ket{+}$ corresponding to $\theta=0,\pi/4$ respectively, in both the reciprocal ($\phi = 0$) and maximally non-reciprocal ($\phi = -\pi/2$) cases. On this figure, we compare the results given by solving numerically Eq.~\eqref{eq:GGEeqNum} and the exact simulation of the model via a stochastic quantum trajectory algorithm for a finite system of $L=14$ sites. The calculation is performed with the open-source python-framework QuTiP~\cite{Johansson2012,Johansson2013}.  Details on how to take care of the finite size of the system and of its boundary conditions can be found in App.~\ref{ap:Ap3}.
We obtain a good agreement between the two methods.

In all simulations presented in Fig.~\ref{fig:rhok}, $\varrho(k)$ tends to develop a peak around the wavevector
\begin{equation}
    \kbar := \pi - \phi
\end{equation}
 at late time; this is compatible with the presence of long-lived modes around $\kbar$ as predicted in Ref.~\cite{Begg2024}. The latter can be understood by noticing that, at late time, the system is almost empty (almost all the spins are down in the spin language), meaning that $n_j \simeq 0, \ \forall j$; thus, we can approximate the string operators as $(-1)^{N_{(-\infty,j-1)}} \simeq 1$ and $(-1)^{n_j} \simeq 1$. The jump operators~\eqref{eq:jumps_fermions} are then approximately equal to the free fermionic ones \textit{i.e.}~$L_j \simeq c_j + e^{i\phi} c_{j+1}$; therefore the momentum distribution is approximately governed by the dynamical equation~\eqref{eq:rhok_ff} which features slow-decaying modes around $k = \kbar$.

On the other hand, following this line of reasoning, in some situations the short-time dynamics can be qualitatively approximated by a free-fermion one.
Indeed, when the initial state is $\ket{\Psi_0} = \ket{\Uparrow}$, we see that $\varrho(k)$ first develops a peak around $\tilde{k}=-\phi$ before developing one around $\kbar$. At short time,  we can appropriate  $(-1)^{n_j} \simeq -1$ and the string operator $(-1)^{N_{(-\infty,j-1)}}$ reduces to an irrelevant scalar number. As a consequence, the jump operators~\eqref{eq:jumps_fermions} become $L_j \simeq c_j - e^{i\phi} c_{j+1} = c_j + e^{i\left( \phi + \pi \right)} c_{j+1}$, and the system behaves similarly to the free-fermionic model~\eqref{eq:ff} with a non-reciprocity angle shifted by $\pi$ \textit{i.e.} $\phi \rightarrow \phi + \pi$. Thus, in the short-time regime, we have
$\frac{d}{dt} \varrho(k) \simeq -2 \kappa \left( 1-\cos(k+\phi) \right) \varrho(k)$, and slow-decaying modes accumulate around $k = \tilde{k}$.

We stress that this reasoning, which consists of approximating the jump operators by free-fermionic ones, is rather crude and only provides a qualitative understanding of the emergence of peaks in the momentum distribution around $k=-\phi$ and $k=\pi-\phi$. In particular, it cannot explain why the late-time decay of the magnetization $n$ displays apparent exotic power-law exponents, because Eq.~\eqref{eq:n_ff} necessarily leads to $n\sim t^{-1/2}$ at late times, regardless of the initial momentum distribution $\vr_0(k)$, as long as $\vr_0(\kbar)\neq0$. This suggests that terms quadratic in the momentum distribution in Eq.~\eqref{eq:GGEeq} cannot be neglected even at late times.

%%%%%%---FIG----%%%%%%
\begin{figure}[t]
\centering
 \includegraphics[width=0.9\columnwidth]{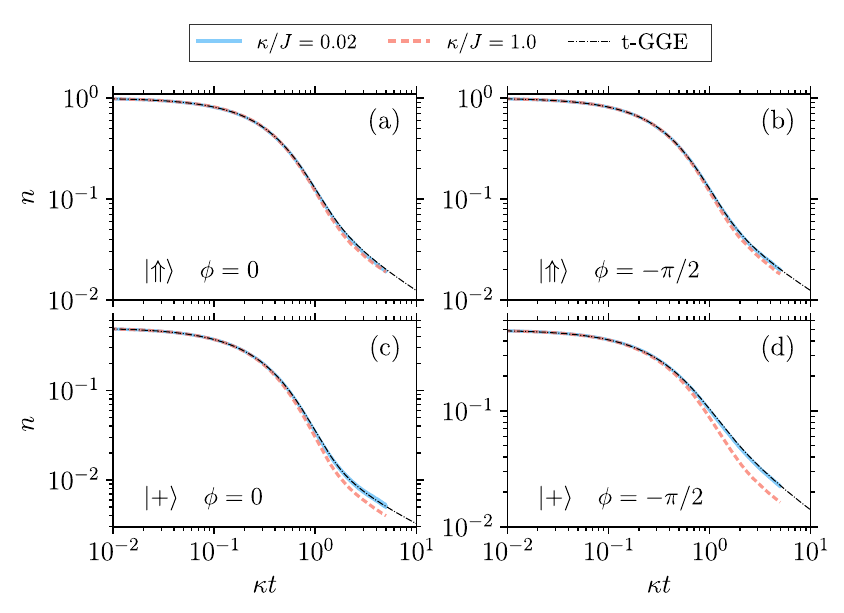}
 \caption{Short time evolution of the local magnetization $n$, for the two initial states $\ket{\Uparrow}$ and $\ket{+}$, in both the reciprocal ($\phi = 0$) and maximally non-reciprocal ($\phi = -\pi/2$) cases. The colored curves are obtained via a stochastic quantum trajectory algorithm with 5000 trajectories for $L=14$ lattice sites, with periodic boundary conditions. The parameters are $J=1$, $\kappa=0.02$ (continuous blue) or $\kappa=1$ (dashed red). The dash-dotted black curves are obtained by solving  Eq.~\eqref{eq:GGEeqNum} and integrating the resulting momentum distribution numerically.}
\label{fig:nQT}
\end{figure}
%%%%%%%%%%%%%%%%%%%%

%%%%%%%------FIG---------%%%%%%%
\begin{figure}[t]
\centering
 \includegraphics[width=1\columnwidth]{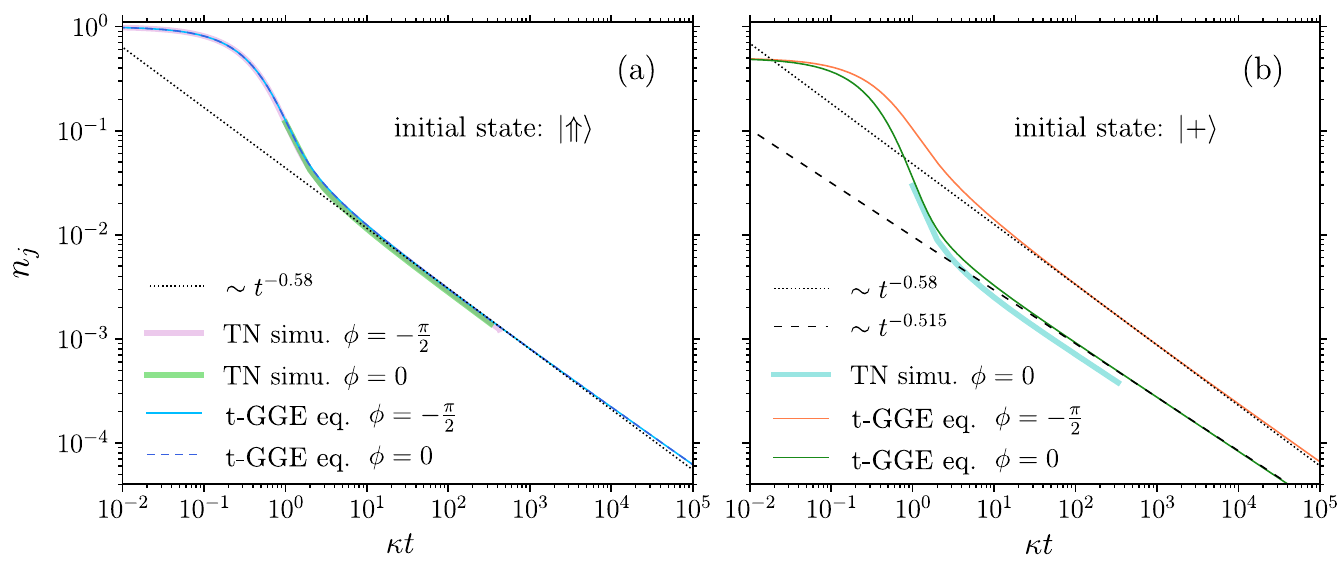}
 \caption{Time evolution of the local magnetization $n_j$ defined in Eq.~\eqref{Eq:n:Spin}, for the two initial states $\ket{\Uparrow}$ and $\ket{+}$, in both the reciprocal ($\phi = 0$) and maximally non-reciprocal ($\phi = -\pi/2$) cases. The wide lines are the results found in Ref.~\cite{Begg2024} via tensor-network (TN) simulations for open boundary conditions and $J/\kappa = 1$ (courtesy of the authors S.E. Begg and R. Hanai for sharing the data). For the purple curve on panel (a), the total number of lattice sites is $L=500$ and the considered site $j=450$; for both the green curve of panel (a) and the blue curve of panel (b),  $L=250$ and $j=L/2$. The thin lines are obtained via the t-GGE 
 equation~\eqref{eq:GGEeqNum}. The data obtained by TN simulations were already sketched in Fig.~\ref{fig:n_sketch}.
 }
\label{fig:n}
\end{figure}
%%%%%%%-----------------%%%%%%%

Finally we note that, if the initial state is $\ket{\Uparrow}$, then the initial momentum distribution $\varrho_0(k)=1$ is uniform; thus varying the non-reciprocity angle $\phi$ amounts simply to a global translation in momentum space (see Figs.~\ref{fig:rhok}~(a) and~(b)). On the contrary, if the initial state is $\ket{+}$, which corresponds to the initial momentum distribution $\varrho_0(k) = \left( 1+ \cos(k) \right)/2$, then varying $\phi$ has a non-trivial effect on $\varrho(k,t)$ (see Figs.~\ref{fig:rhok}~(c) and~(d)).

%%%%%%%------FIG---------%%%%%%%
\begin{figure}[t]
\centering
 \includegraphics[width=0.85\columnwidth]{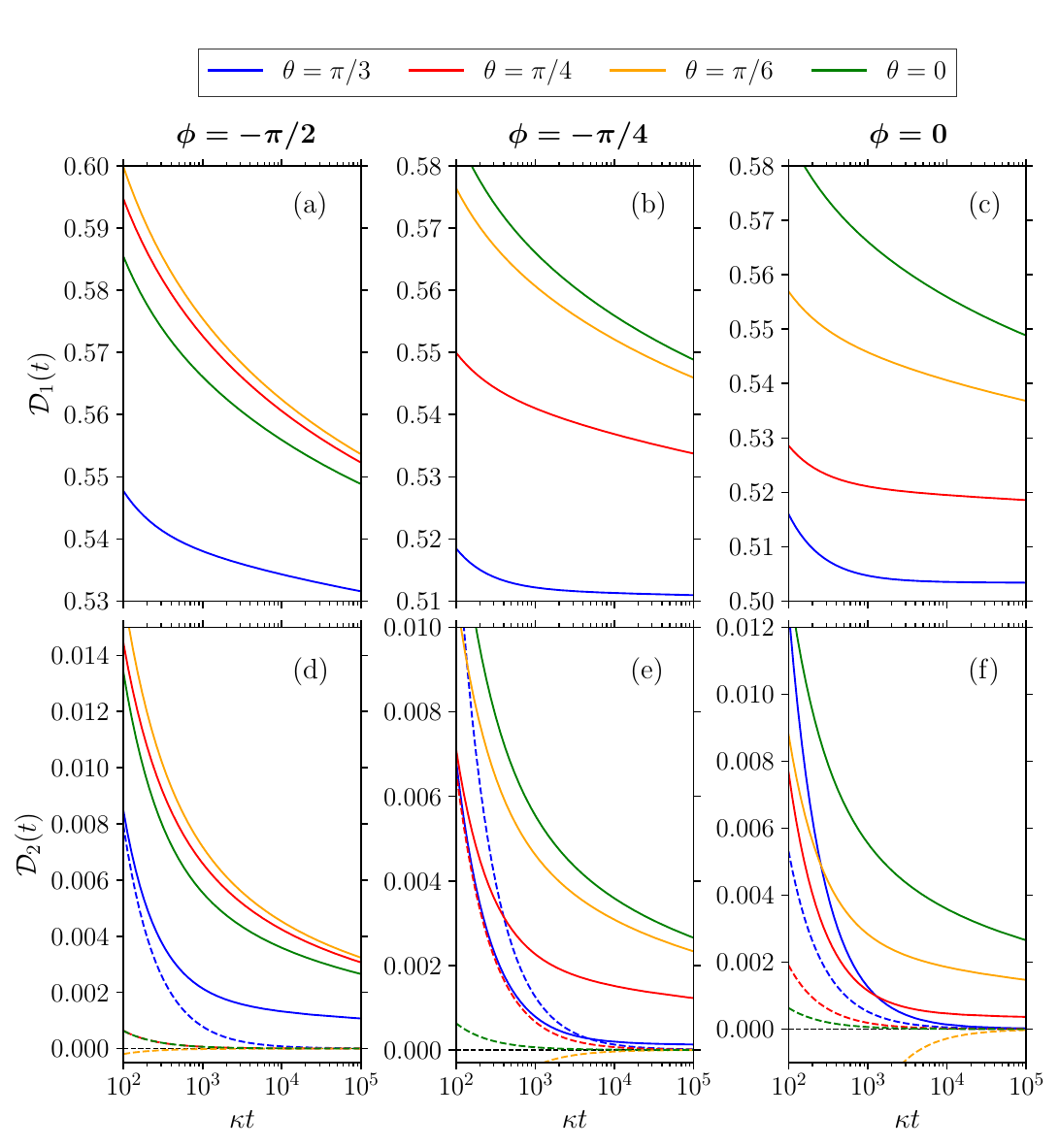}
 \caption{First and second logarithmic derivatives of $n$, for various initial states parameterized by~$\theta$ and various strength of non-reciprocity~$\phi$. The continuous curves are obtained via the t-GGE approach in Eq.~\eqref{eq:GGEeqNum} describing the dynamics of the spin model.
 The dashed curves in panels (d-f) are obtained via the dynamical Eqs.~\eqref{eq:rhok_ff} describing the time evolution of the free fermionic model~\eqref{eq:ff}.}
\label{fig:log_der}
\end{figure}
%%%%%%%-----------------%%%%%%%

%%%%%%%%%%%%%%----------------------------------------------------------------------------------%%%%%%%%%%%%%%
\section{Time decay of the magnetization: numerical observations}
\label{Sec:Magnetization}
%%%%%%%%%%%%%%----------------------------------------------------------------------------------%%%%%%%%%%%%%%

\subsection{Comparison between the t-GGE results and microscopic simulations} \label{sec:compGGE_TN}

We focus on the dynamics of the local magnetization $n$, defined in Eq.~\eqref{Eq:Def:Fermi:Dens}; we could not find a closed-form expression for its dynamics implied exactly by the momentum-distribution dynamics in Eq.~\eqref{eq:GGEeq}. Hence, we compute $n(t)$ by numerical integration of the numerical solution for $\varrho(k,t)$ discussed in Sec.~\ref{SubSec:T:rhok}.   
We check that this method is consistent with exact simulations of a small system of $L=14$ sites using quantum trajectories, for times up to $\kappa t=5$, as shown in Fig.~\ref{fig:nQT}. We compute $n$ with quantum trajectories for both $\kappa/J=0.02$ and $\kappa/J=1$. As anticipated, the t-GGE equations~\eqref{eq:GGEeqNum} give accurate results in the weakly dissipative regime $\kappa \ll J$. Moreover, it also provides surprisingly good predictions in the case $\kappa/J=1$ when the initial state is $\ket{\Uparrow}$ (see Figs.~\ref{fig:nQT}~(a) and~(b)).
This is not the case for $\varrho(k)$: however, errors are averaged out by the integral in Eq.~\eqref{Eq:Def:Fermi:Dens}.

Let us now focus on the tensor-network simulations of $n$ presented in Ref.~\cite{Begg2024}, which have been performed for significantly large systems composed of $L=250$ or $500$ sites, whose results have been already discussed in Sec.~\ref{sec:resultsBH}. 
We now compare those results with those obtained with our t-GGE method~\eqref{eq:GGEeqNum}, which are displayed in Fig.~\ref{fig:n} for the four cases obtained by considering both $\phi= 0,-\pi/2$ and initializing the system in both states $\ket{\Psi_0} = \ket{\Uparrow},\ket{+}$.
When the comparison is possible, the agreement is very good, especially if we consider that the tensor-network simulations are performed in a regime, $\kappa = J$, where our perturbative approach is not expected to work.
More precisely, we see that our results and the simulations of Ref.~\cite{Begg2024} superimpose when $\ket{\Psi_0} = \ket{\Uparrow}$, see Fig.~\ref{fig:n}~(a), while a vertical shift is present when $\ket{\Psi_0} = \ket{+}$, see Fig.~\ref{fig:n}~(b).
We also mention that, from the discussion of Sec.~\ref{sec:rhok}, it is clear that, for $\ket{\Psi_0} = \ket{\Uparrow}$, the dynamics of $n$ is completely insensitive to the value of $\phi$, since varying $\phi$ results only in a global shift of the distribution $\varrho(k)$ in momentum space.

It could seem surprising at first sight that the same apparent power-law exponents emerge at late times for very different regimes of dissipation strength: $\kappa \ll J$ with the t-GGE approach and $\kappa = J$ for the tensor-network simulations. However, we believe that this could generally occur when the Lindbladian gap closes in the thermodynamic limit. For instance, in bosonic gases trapped in one spatial dimension and subjected to two-body losses, after a non-universal transient regime, there is a power-law exponent emerging at late time in the decay of the number of particles that depends on the initial state but is completely independent of the Lindbladian parameters (in this specific case the elastic interaction strength and the loss rate)~\cite{Marche2024,hosseinabadi2026}.

\subsection{Late-time decay} \label{sec:num_obs_late_time}

We now focus specifically on the behavior of the magnetization at late times.
Fig.~\ref{fig:n} shows that in all situations, for times $50 \lesssim  \kappa t \lesssim 10^{4}$, the t-GGE approach yields apparent power-law decays that are compatible with those of the tensor-network data.
Specifically, we observe $n \sim t^{-\chi}$ where $\chi=0.58$ for $\ket{\Psi_0} = \ket{\Uparrow}$ and $\phi = 0,-\pi/2$, while $\chi=0.515$ for $\ket{\Psi_0} = \ket{+}$ and $\phi = 0$. We also report that $\chi=0.58$ for $\ket{\Psi_0} = \ket{+}$ and $\phi = -\pi/2$, a situation that is not considered in Ref.~\cite{Begg2024}.

The t-GGE methods allows us to access time scales much larger than those reachable via tensor-network simulations, such as those employed in Ref.~\cite{Begg2024}, which do not exceed $ \kappa t \sim 10^3$.
By pushing our numerics to times as long as $\kappa t \sim 10^5$ we can inspect whether the identified scalings $t^{-\chi}$ persist at larger times; the convergence of these numerics have been carefully checked (see App.~\ref{ap:Ap2}).
When the initial state is $\ket{\Uparrow}$, we observe a small but significant deviation from a pure power law decay in the time window  $10^4 \lesssim  \kappa t \lesssim 10^{5}$, as it is visible in Fig.~\ref{fig:n}~(a). To investigate this behavior, we analyze the first and second logarithmic derivatives of $n(\kappa t)$ with respect to $\log( \kappa t)$, defined respectively as
\begin{equation}
 \mathcal{D}_1(t) := - \frac{d \log (n)}{d \log (\kappa t)} \quad \text{ and } \quad \mathcal{D}_2(t) := \frac{d^2 \log (n)}{(d \log (\kappa t))^2}. \label{eq:def_D1_D2}
\end{equation}
For a pure algebraic decay $n ~ \sim (\kappa t)^{-\chi}$, the first logarithmic derivative is time-independent and equal to the critical exponent $\chi$, while the second logarithmic derivative vanishes: $\mathcal{D}_1(t) = \chi$ and $\mathcal{D}_2(t) = 0$.

The numerical results are presented in Fig.~\ref{fig:log_der}
for several values of the non-reciprocity parameter $\phi = -\pi/2$, $-\pi/4$, $0$ and for different initial states parametrized by the angles $\theta = \pi/3$, $\pi/4$, $\pi/6,0$.
For most values of the angles $\theta$ and $\phi$, we observe that $\mathcal{D}_1(t)$ continues to decay without approaching a stationary value, and that \(\mathcal{D}_2(t)\) remains finite over the entire time window accessible with the t-GGE method.  We thus conclude that $n(t)$ is not converging towards a purely algebraic decay behavior within the probed time scales; the only exception to this seems to be the case $\theta=\pi/3$ and $\phi = 0$.
In contrast, for the same time interval, we see that the free fermionic system displays a power-law behavior since $\mathcal{D}_2(t)$ reaches zero (see dashed lines in Figs.~\ref{fig:log_der} (d) (e) (f)).

On the basis of these numerical observations alone, it remains unclear whether (i) the magnetization in the spin system eventually crosses over to a true algebraic decay at extremely late times, or (ii) a deviation from power-law behavior persists asymptotically, so that the late-time relaxation of the spin system is qualitatively different from that of free fermions.
In Refs.~\cite{Perfetto_2023_2,Lehr2025}, the authors found a similar phenomenology where a possible correction to purely algebraic decay is revealed numerically by computing a quantity similar to $\mathcal{D}_1$.
Moreover, in a previous work, we pointed out that in some lattice models the late-time decay could display a logarithmic correction that could be extremely difficult to fit~\cite{Marche2024}; at this point in the discussion, it is still to be understood whether this is also what happens here.

%%%%%%%%%%%%%%%%%%%%%%%%%%%%%%%%%%%%%%%%%%%%%%%%%%%%%%%%%%%%%%%%%%%%%%%%%%%%%%%%%%%%%%%%%%%%%%%%%%%%%%%%%%%%%%%
%%%%%%%%%%%%%%%%%%%%%%%%%%%%%%%%%%%%%%%%%%%%%%%%%%%%%%%%%%%%%%%%%%%%%%%%%%%%%%%%%%%%%%%%%%%%%%%%%%%%%%%%%%%%%%%
\section{Hamiltonian magnetization current}
\label{Sec:Current}

%%%%%%%------FIG---------%%%%%%%
\begin{figure}[t]
 \includegraphics[width=\columnwidth]{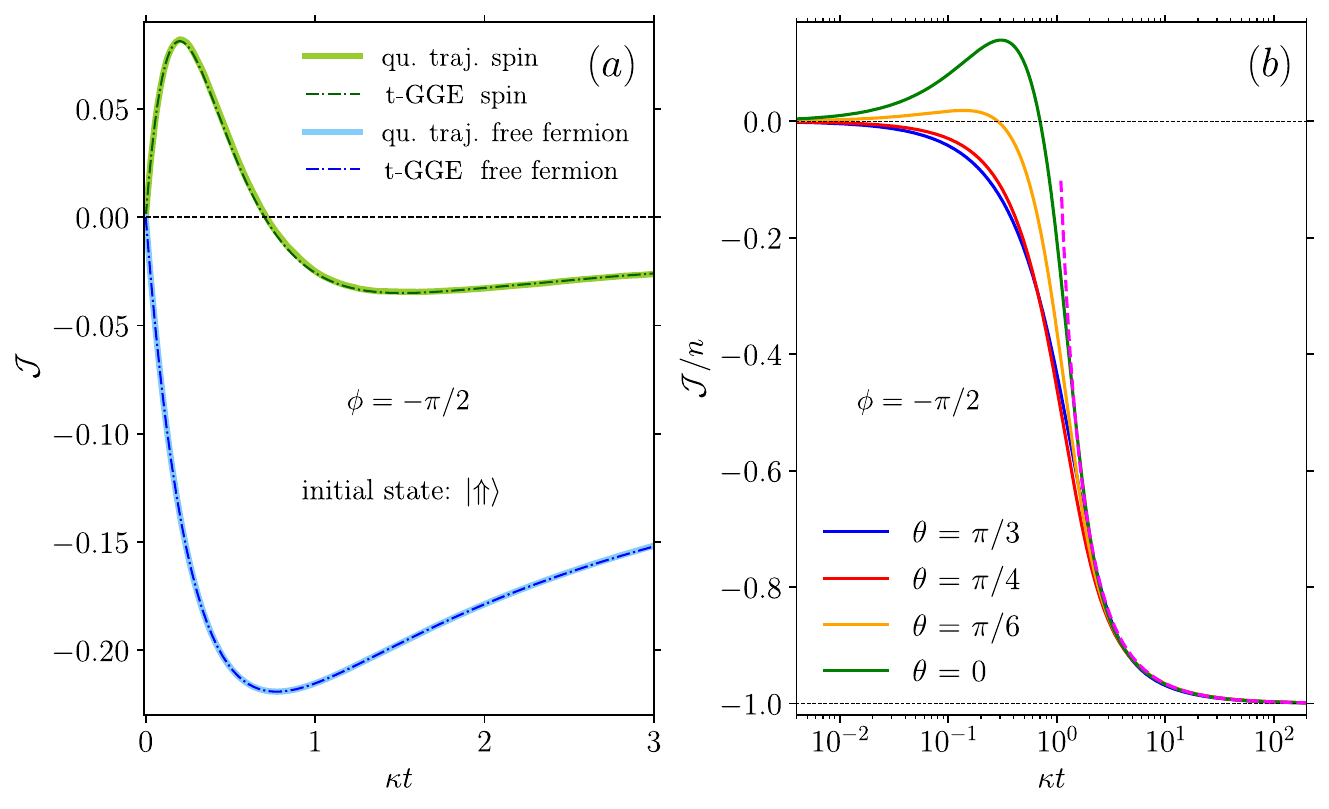}
 \caption{(a) Time evolution of the magnetization current for the spin system described by Eqs.~\eqref{eq:model} (in green), and for the free fermionic system described by Eqs.~\eqref{eq:ff} (in blue).
 The initial state is $\ket{\Uparrow}$ and the non-reciprocity is maximal, i.e.~$\phi = - \pi/2$. The dash-dotted curves are obtained by solving numerically the t-GGE equations~\eqref{eq:GGEeqNum} and the dynamical equations~\eqref{eq:rhok_ff}, respectively, for the integral expression of the current $\mathcal{J}  = J \int dk \sin(k)\varrho(k)/2\pi$.  The continuous curves are obtained via stochastic quantum trajectories averaging over $5000$ trajectories, with periodic boundary conditions and $\kappa/J =0.01$. For the free fermions, we use $L=10$ lattice sites; for the spin system, we use $L=14$ lattice sites. (b) Time evolution of the ratio $\mathcal{J}/n$ for $\phi = - \pi/2$ and for various initial states parametrized by the angle $\theta$, in the case of the spin system. The continuous curves are obtained via the t-GGE equations~\eqref{eq:GGEeqNum}. The dashed magenta curve is $f(\kappa t) = \sin(\phi) \left( 1- \sigma(\kappa t)^2/2\right)$ where $\sigma(\kappa t)$ is obtained by fitting the momentum distribution at time $t$ with the gaussian~\eqref{eq:gauss}, in the case $\theta=0$.  For both panels (a) and (b), $J=1$. }
\label{fig:curr}
\end{figure}
%%%%%%%-----------------%%%%%%%

In this section, we discuss several properties of the Hamiltonian magnetization current density, defined as
\begin{equation}
 \mathcal{J} := \langle \frac{i J}{2 } \left( S_{j+1}^+ S_j^- - S_j^+ S_{j+1}^- \right) \rangle  =  \langle \frac{i J}{2 } \left( c_{j+1}^\dag c_j - c_j^\dag c_{j+1} \right) \rangle  = J \int_0^{2 \pi} \frac{dk}{2 \pi} \sin(k) \varrho(k).
\end{equation}
We recall that in our problem the latter is invariant by translation in real space, hence the absence of a dependence on $j$ in the final expression.
Note that this is not the operator that satisfies the continuity equation for the magnetization under the dissipative dynamics, but only under the Hamiltonian one~\cite{Keck2018, Soares2025}.

We first remark that, depending on the strength of the non-reciprocity $\phi \neq 0$ and the initial state, a sign reversal of the current can occur dynamically for the spin system, while this phenomenon is totally absent for the analogous free fermionic system. This behavior occurs for instance when $\ket{\Psi_0} = \ket{\Uparrow}$ and $\phi = -\pi/2$ as shown in Fig.~\ref{fig:curr}~(a):  we indeed observe a reversal of the current sign for the spin system, while  the current remains negative at all times for the free fermions. As shown in Fig.~\ref{fig:rhok}~(b), the momentum distribution $\varrho(k)$ develops a peak at momenta between $0$ and $\pi$ at short times, which implies $\mathcal{J}>0$. At later times, however, a second peak necessarily emerges at momenta between $\pi$ and $2\pi$, resulting in $\mathcal{J}<0$.
The different behavior at high and low fillings can be understood with the simple arguments proposed in Sec.~\ref{SubSec:T:rhok} and mapping the dynamics to two different models in the two cases.

Another straightforward consequence of the fact that the slow–decaying modes are located around $\kbar = \pi - \phi$ is that the ratio $\mathcal{J}/n$ becomes constant and equal to $J \sin(\phi)$ in the limit $t \rightarrow \infty$, as shown in Fig.~\ref{fig:curr}~(b). Indeed, at late times the distribution $\rho(k)$ is sharply peaked around $\kbar$, which allows us to perform a saddle-point approximation: $\mathcal{J} = J \int_0^{2 \pi} dk \varrho(k) \sin(k)/2 \pi  \simeq J \sin(\kbar)  \int_0^{2 \pi} dk \varrho(k)/2 \pi = J \sin(\kbar) n$. Similarly, the ratio between density of energy $\varepsilon := -J\int_0^{2 \pi} dk \cos(k) \varrho(k)/2\pi$ and $n$ tends to $J \cos(\phi)$ when $t \rightarrow \infty$;
the ratio $\mathcal J/ \varepsilon$ tends to $\tan(\phi)$.

We note that the way $\mathcal{J}/n$ approaches its stationary value gives information about the width of the momentum distribution. For instance, in the case $\ket{\Psi_0} = \ket{\Uparrow}$, at late time the momentum distribution is well-approximated by a gaussian of the form
\begin{equation}
    \varrho(k,t) = \mathcal{A}(\kappa t) \exp(-(k-\kbar)^2/2 \sigma(\kappa t)^2) \label{eq:gauss},
\end{equation}
where $\mathcal A(\kappa t)$ and $\sigma(\kappa t)$ are time-dependent functions.
Thus, by Taylor expanding $\sin(k)$ up to the $O(k^2)$ order, we find after a few algebraic manipulations that
\begin{equation}
    \mathcal{J}/n \simeq \sin(\phi) \left( 1 - \sigma(\kappa t)^2/2\right).
\end{equation}
The latter identity is checked numerically via the t-GGE equation~\eqref{eq:GGEeqNum}, see Fig.~\ref{fig:curr}~(b).
%%%%%%%%%%%%

%%%%%%%%%%%%%%%%%%%%%%%%%%%%%%%%%%%%%%%%%%%%%%%%%%%%%%%%%%%%%%%%%%%%%%%%%%%%%%%%%%%%%%%%%%%%%%%%%%%%%%%%%%%%%%%
%%%%%%%%%%%%%%%%%%%%%%%%%%%%%%%%%%%%%%%%%%%%%%%%%%%%%%%%%%%%%%%%%%%%%%%%%%%%%%%%%%%%%%%%%%%%%%%%%%%%%%%%%%%%%%%
\section{Analytical characterization of the magnetization late-time decay}
\label{Sec:AnalyticalCharacterization}

In this section, we discuss the possibility of characterizing the magnetization late-time decay $n(t)$ in an analytical way, thus pushing further the purely numerical analysis done in Sec.~\ref{sec:num_obs_late_time}. 
During the resubmission of the present article, a closely related work~\cite{daraban2026} appeared, in which
the authors studied the spin model~\eqref{eq:model}, in the case $\phi=\pi$, for an initially uniform momentum distribution, i.e.~$\vr_0(k)$ constant. 
The authors identify the universal dynamics of the model, including the decay of the zero mode, and the universal scaling form of the momentum distribution, including the Gaussian profile and the algebraic tails. Additionally,
they found that the asymptotic behavior of $n(t)$, in the limit $t \rightarrow \infty$, is a logarithmically corrected decay of the form $\sim (\kappa t)^{-1/2} \log(\kappa t)^{-1}$.
This is the first article where a logarithmic correction is analytically identified and proved in a many-body loss process. Moreover, in an appendix, the authors identify corrections to this leading late-time behavior, including a $\log \log t$ correction, which allow to extend the prediction for the magnetization dynamics to shorter times.

Stimulated and inspired by this new advance, we present in this section our approach to analytically characterize the late-time behavior of $n(t)$. We derive a closed differential equation for the magnetization which captures in a satisfactory manner the intermediate-time decay of the magnetization.
To do so, we study the quantity 
\begin{equation}
 w_r(t) :=  \int_0^{ 2 \pi} \frac{dk}{2 \pi} \vr(k,t) \cos((k-\bar{k})r), \quad \forall r \in \mathbb{Z}, \label{eq:def_w_r}
\end{equation}
which turns out to be relevant since at late time $\vr(k)$ is peaked around $\bar{k} = \pi - \phi$, as discussed in Sec.~\ref{SubSec:T:rhok}. Here, $w_r(t)$ can be interpreted as a kind of modified Fourier transform of $\vr(k)$, and it is worth noting that $n(t) = w_0(t)$.

\subsection{Free fermions}

In order to get a better grasp of the quantity $w_r(t)$, we first discuss the free fermionic case. We can show, using Eq.~\eqref{eq:rhok_ff} and~\eqref{eq:def_w_r}, that, at late times, the profile takes the form
\begin{equation}
 w_r(t) \simeq w_0(t) f_{\rm FF}\left(\frac{r}{2\sqrt{\kappa t}} \right),  \ \text{with} \ f_{\rm FF}(x) = \begin{cases}
     e^{-x^2} \left( 1 - 2 x^2\right),  &\text{if } \phi=0 \text{ and } \theta \in (0 ,\frac{\pi}{2}),\\
     e^{-x^2}, & \text{otherwise};
 \end{cases} \label{eq:w_r_gaussian_ff}
\end{equation}
see App.~\ref{ap:w_r_ff} for details. In particular, we remark that it exhibits diffusive broadening, since its typical width scales as $1/\sqrt{\kappa t}$.

\subsection{Spin model in Eq.~\eqref{eq:model}} \label{sec:w_r_spin}

%%%%%%%------FIG---------%%%%%%%
\begin{figure}[t]
\centering
 \includegraphics[width=0.8\columnwidth]{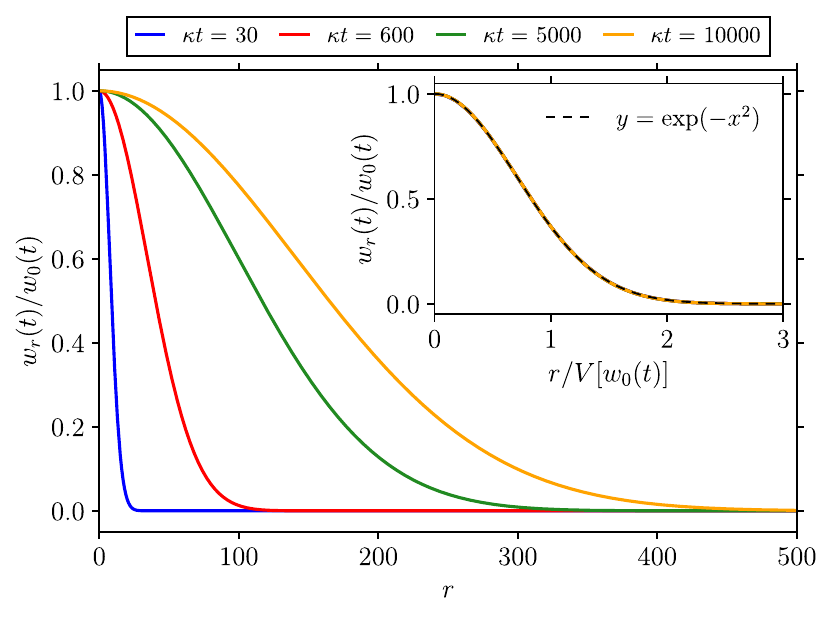}
 \caption{Profiles of $w_r(t)$, obtained by solving numerically the t-GGE equations~\eqref{eq:GGEeqNum}, for different fixed times $\kappa t \in \{ 30, 600,5000,10000\}$. The strength of the non-reciprocity is $\phi=-\pi/4$ and the initial state is given by~$\theta = \pi/3$. In the inset, the curves are rescaled according to Eqs.~\eqref{eq:Ansatz} and~\eqref{eq:delta_of_u0}, and collapse perfectly on a Gaussian. Here, we use $A = 81.36$.}
\label{fig:w_n}
\end{figure}
%%%%%%%-----------------%%%%%%%

We now turn to the spin model~\eqref{eq:model}. In Fig.~\ref{fig:w_n}, we show the associated profile $w_r(t)$, for different fixed times, computed numerically from the t-GGE equation~\eqref{eq:GGEeqNum}. We observe that it takes a gaussian shape for any value of $\phi$ and $\theta$.
Together with Eq.~\eqref{eq:w_r_gaussian_ff}, these observations suggest to postulate an ansatz of the form
\begin{equation}
 w_r(t) = w_0(t) f\left(\frac{r}{\ww(t)}\right), \quad  \text{with} \quad f(x) = e^{-x^2}. \label{eq:Ansatz}
\end{equation}
Here, $\ww(t)>0$ is the typical width of the gaussian profile at time~$t$; it increases with time, as shown in Fig.~\ref{fig:w_n}, but, crucially we do not assume a diffusive scaling.

In the ansatz~\eqref{eq:Ansatz}, the respective time evolutions of both the height $w_0(t)$ and the width $\ww(t)$ are \textit{a priori} unknown. In order to find their dynamics, we use the following two equations relating the different $\{ w_r \}$:
\begin{subequations} \label{eq:rel_w_r}
    \begin{equation}
  \dot{w}_0  = 2 \kappa \left( w_1 -  w_0 \right), \label{eq:dw0}
\end{equation}
\begin{equation}
   \dot{M} = 2 \dot{w}_0 K - 4 \kappa w_0^2, \quad \text{where} \quad M = \sum_{r=-\infty}^{\infty} w_r, \quad \text{and} \quad K = \sum_{r=-\infty}^{\infty} |r| w_r \label{eq:dM}.
\end{equation} 
\end{subequations}
Eqs.~\eqref{eq:rel_w_r} can be directly derived from the t-GGE equations~\eqref{eq:GGEeqNum}, without any further approximations; see App.~\ref{ap:w_r}. In particular we do not need to assume the ansatz~\eqref{eq:Ansatz} to show them. We note that Eqs.~\eqref{eq:rel_w_r}, expressed in terms of the Fourier coefficients $w_r$, are equivalent to equations already derived in Ref.~\cite{daraban2026} in the reciprocal case $\phi=\pi$.

\begin{figure}[h!]
\centering
 \includegraphics[width=0.9\columnwidth]{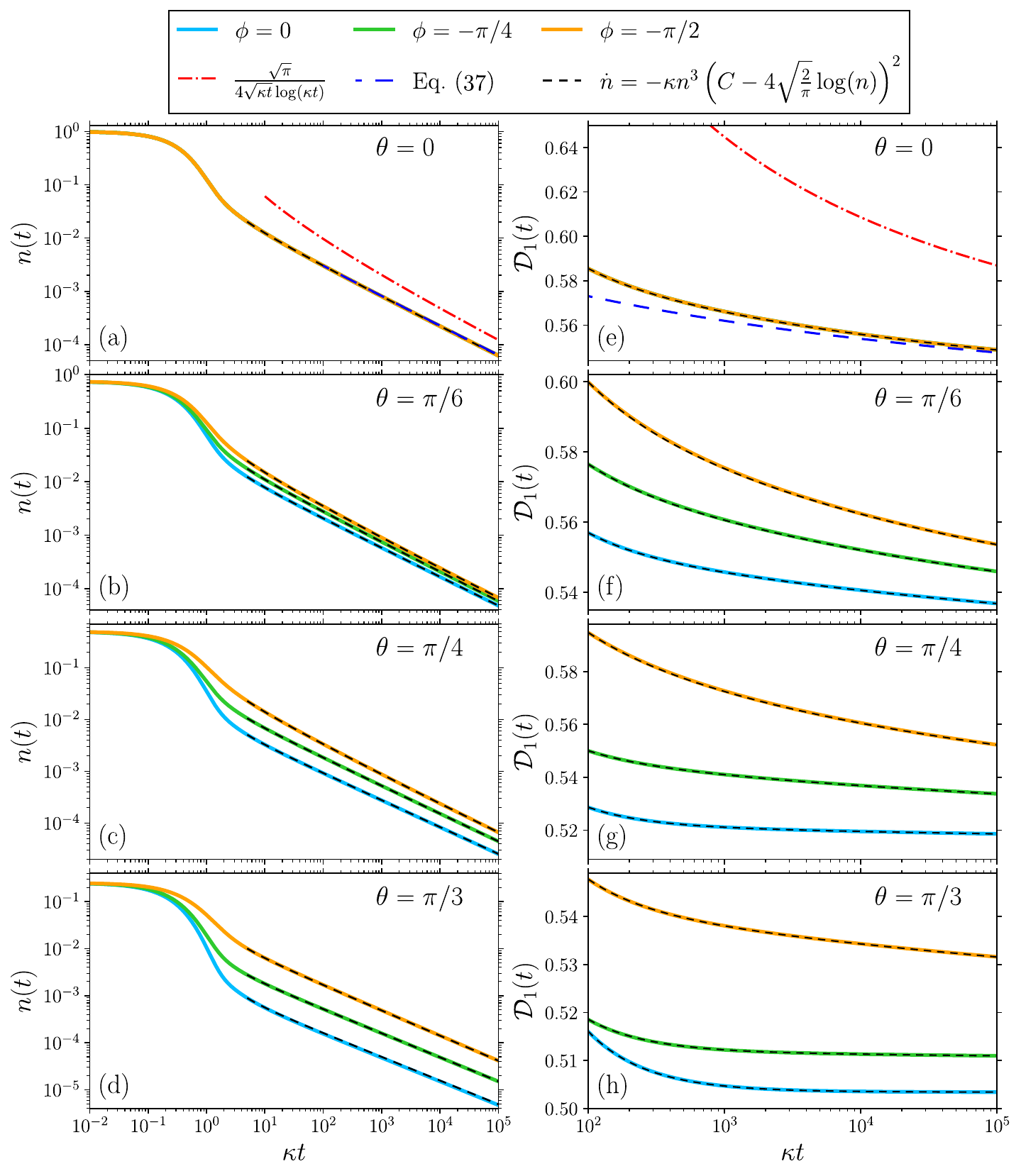}
 \caption{Time evolutions of the local magnetization~$n(t)$ (panels (a--d)) and its first logarithmic derivative~$\mathcal{D}_1(t)$  (panels (e--h)), for various initial states parametrized by~$\theta$ and various strength of the non-reciprocity~$\phi$. The continuous colored curves are obtained via the t-GGE equations~\eqref{eq:GGEeqNum}. The dashed black curves are obtained by solving numerically the closed equation~\eqref{eq:dn_ana}, for~$t>t_0$; the values of the initial condition~$n(t_0)$ and the constant $C$ are chosen such that the curves $n(t)$ obtained via Eqs.~\eqref{eq:GGEeqNum} and~\eqref{eq:dn_ana} match at~$t = t_0$. We choose $\kappa t_0 = 5$ for the panels (a--d) and~$\kappa t_0 = 100$ for the panels (e--h). On the panels~(a) and~(e), the dash dotted red line corresponds to the asymptotic form~\eqref{eq:n_asympotic}, first derived in Ref.~\cite{daraban2026}; the blue dashed line corresponds to Eq.~\eqref{eq:n_asympotic2} with $\kappa t_0=100$, the value of $n(t_0)$ is chosen such that the curves $n(t)$ obtained via Eqs.~\eqref{eq:GGEeqNum} and~\eqref{eq:n_asympotic2} match at~$ \kappa t_0 = 100$.
 }
\label{fig:dn}
\end{figure}

By substituting Eq.~\eqref{eq:Ansatz} into Eqs.~\eqref{eq:rel_w_r}, and assuming that $\ww(t) \gg 1$, we obtain the closed differential system
\begin{subequations} \label{eq:u0_delta}
\begin{equation}
\dot{w}_0 \simeq - 2 \kappa \frac{w_0}{\ww^2}, \label{eq:du0_of_delta}
\end{equation}
\begin{equation}
 \ \dot{\ww} \simeq  \kappa \left(\frac{2}{\ww} - \frac{8}{\sqrt{\pi}} w_0\right).
\end{equation}
\end{subequations}
From this, we find the following relation between the width and the height of the gaussian profile:
\begin{equation}
 \ww = w_0^{-1} \left( A - \frac{4}{\sqrt{\pi}} \log(w_0)\right)^{-1} = V[w_0], \label{eq:delta_of_u0}
\end{equation}
where $A$ is a constant of integration. Eqs.~\eqref{eq:u0_delta} and~\eqref{eq:delta_of_u0} are proved in App.~\ref{ap:w_r}. In the inset of Fig.~\ref{fig:w_n}, we show the collapse of the rescaled profiles $w_r(t)$ according to Eqs.~\eqref{eq:Ansatz} and~\eqref{eq:delta_of_u0}.
Finally, substituting Eq.~\eqref{eq:delta_of_u0} into Eq.~\eqref{eq:du0_of_delta}, and using the fact that $w_0(t) = n(t)$, we obtain
\begin{equation}
 \dot{n} = - \kappa n^3 \left( C - 4 \sqrt{\frac{2}{\pi}} \log(n) \right)^2. \label{eq:dn_ana}
\end{equation}
Here, $C = \sqrt{2}A$ is a constant which depends on both $\theta$ and $\phi$. Equation~\eqref{eq:dn_ana} is the second main result of this article. Since this equation does not admit an explicit solution in terms of elementary functions, we solve it numerically. 
In Fig.~\ref{fig:dn}~(a--d), we compare the decays of magnetization~$n(t)$ obtained respectively via the t-GGE equations~\eqref{eq:GGEeqNum} and the Eq.~\eqref{eq:dn_ana}, for various parameters $\theta$ and $\phi$; we observe a very good agreement. To push even further the test of this agreement, we also compare the first logarithmic derivatives of the magnetization $\mathcal{D}_1$, defined as in Eq.~\eqref{eq:def_D1_D2},
which we compute numerically and plot in Fig.~\ref{fig:dn}~(e--h); we again observe a perfect agreement of the data.

A few remarks can be formulated concerning Eq.~\eqref{eq:dn_ana}. The $\log(n)$ contribution on the right hand side of this equation is solely responsible for the deviation from a purely algebraic behavior; a rate equation of the form $ \dot{n} = - \kappa n^3$ would imply a decay $n \sim t^{-1/2}$. Moreover, in the free-fermionic case,  we recall that, depending on the values of the angles $\theta$ and $\phi$, two different power-law decays can emerge, $n\sim t^{-1/2}$ or $n \sim t^{-3/2}$, however we stress that for the spin model~\eqref{eq:model}, Eq.~\eqref{eq:dn_ana} is valid for any values of $\theta$ and $\phi$: in that sense Eq.~\eqref{eq:dn_ana} is universal.

%%%---FIG----%%%%
%%%---------%%%%

The asymptotic solution of Eq.~\eqref{eq:dn_ana}, in the limit $t \rightarrow \infty$, is
\begin{equation}
 n(t) \simeq  \frac{\sqrt{\pi}}{4 \sqrt{\kappa t} \log(\kappa t)} \label{eq:n_asympotic};
\end{equation}
see App.~\ref{ap:Link_MM} for a proof. Eq.~\eqref{eq:n_asympotic} is precisely the result presented in Ref.~\cite{daraban2026}, thus our approach is fully compatible with this previous finding;
the red dot-dashed curves presented in Fig.~\ref{fig:dn} are Eq.~\eqref{eq:n_asympotic}.
For completeness, we mention that, according to the supplemental material of Ref.~\cite{daraban2026}, an expression of the magnetization decay retaining subleading corrections with respect to the asymptotic solution~\eqref{eq:n_asympotic}, is 
\begin{equation}
    n(t) \simeq  - \frac{\sqrt{\pi}}{4 \sqrt{ \kappa t} W_{-1} \left(\frac{-\sqrt{\pi\kappa t_0}}{4 \kappa tn(t_0) } \exp(\frac{- \sqrt{\pi}}{4 \sqrt{\kappa t_0} n(t_0)}) \right)},  \label{eq:n_asympotic2}
\end{equation}
where $W_{-1}$ is the Lambert $W$ function for the $-1$ branch and $t_0 \sim \kappa^{-1}$ is a time for which the scaling solution
%~\eqref{eq:scaling_sol_MM} 
has set in; this solution is plotted as a blue dashed line in Fig.~\ref{fig:dn}.

\subsection{Comparison of Eq.~\eqref{eq:dn_ana} with the previous literature}

We believe that Eq.~\eqref{eq:dn_ana} has an advantage with respect to Eqs.~\eqref{eq:n_asympotic} and~\eqref{eq:n_asympotic2} because it describes the dynamics starting from shorter times. Indeed, it appears that the timescales needed to converge towards the asymptotic behavior~\eqref{eq:n_asympotic}  are extremely long and thus inaccessible to state-of-the-art numerical simulations such as those based on tensor-network methods. 
Eq.~\eqref{eq:dn_ana} describes the dynamics for times as short as $\kappa t \simeq 5$; this dynamical equation is valid as soon as the profile $w_r(t)$ takes a gaussian shape and the associated width $\ww(t)$ is large enough.
We have observed that the asymptotic behavior proposed by Eq.~\eqref{eq:n_asympotic} is associated to a diffusive broadening of the Gaussian postulated in Eq.~\eqref{eq:Ansatz}, with $\ww(t) \simeq  2 \sqrt{\kappa t}$. To correctly capture the intermediate times with Eq.~\eqref{eq:dn_ana} one has instead to determine the dynamics of $\ww(t)$ in a self-consistent way.

%%%%%%%%%%%%%%%%%%%%%%%%%%%%%%%%%%%%%%%%%%%%%%%%%%%%%%%%%%%%%%%%%%%%%%%%%%%%%%%%%%%%%%%%%%%%%%%%%%%%%%%%%%%%%%%
%%%%%%%%%%%%%%%%%%%%%%%%%%%%%%%%%%%%%%%%%%%%%%%%%%%%%%%%%%%%%%%%%%%%%%%%%%%%%%%%%%%%%%%%%%%%%%%%%%%%%%%%%%%%%%%
\section{Conclusion and perspectives}
\label{Sec:Conclusions}
%%%%%%%%%%%%%%%%%%%%%%%%%%%%%%%%%%%%%%%%%%%%%%%%%%%%%%%%%%%%%%%%%%%%%%%%%%%%%%%%%%%%%%%%%%%%%%%%%%%%%%%%%%%%%%%
%%%%%%%%%%%%%%%%%%%%%%%%%%%%%%%%%%%%%%%%%%%%%%%%%%%%%%%%%%%%%%%%%%%%%%%%%%%%%%%%%%%%%%%%%%%%%%%%%%%%%%%%%%%%%%%

We have presented a study based on the time-dependent generalized Gibbs ensemble of the dissipative dynamics of an open and non-reciprocal quantum spin chain.
This problem, that has already been the object of several previous publications, is characterized by various peculiar properties, such as the fact that a current is dynamically established only due to the dissipative dynamics~\cite{Keck2018}.
Extensive tensor-network simulations have shown some peculiar properties of the power law that describes the temporal decay of the magnetization of the chain~\cite{Begg2024}.

We believe that the main interest of this approach is to provide a simple theoretical framework for dealing with a Markovian open quantum spin chain, whose coherent dynamics is governed by the XX model; our framework produces controlled and accurate predictions in the limit of small dissipation, while producing data with an almost quantitative agreement also for larger dissipation strengths.
The Eqs.~\eqref{eq:GGEeq} and~\eqref{eq:GGEeqNum} constitute the main result of our work, and, although they do not seem to be amenable to a simple analytical solution, they can be easily solved numerically.
The plot in Fig.~\ref{fig:rhok} shows the main physical message of our work, namely the dynamics of the momentum distribution under this non-reciprocal dynamics.
Non-reciprocity is responsible for the development of a momentum distribution that is not symmetric under the exchange $k \to -k$, and that develops a magnetization current. In Fig.~\ref{fig:curr} we have shown that the Hamiltonian current of the model can display an intriguing switch of direction during the time evolution.

Another problem that we addressed is the apparent power-law decay of the magnetization of the spin chain, which was shown in Ref.~\cite{Begg2024} to be characterized by unusual power-law exponents. Our study can access longer time scales and suggests that these exponents depend on time. 
In order to derive an analytical theory for this behavior, we took great inspiration from the remarkable results in Ref.~\cite{daraban2026}, which appeared after the first version of this article had been submitted, while we were working on the resubmission. 
The authors of Ref.~\cite{daraban2026}
identified the universal dynamical scaling, including the logarithmic decay of the zero mode and the scaling form of the momentum distribution.
In our study, we demonstrated that a Gaussian ansatz for the momentum distribution with a time-dependent width yields a closed equation for the magnetization that reproduces the leading asymptotic behavior and improves the quantitative description at intermediate times.

We believe that our article opens several interesting perspectives.
Unusual power-law exponents have also been reported in numerical studies of other models, for example in Ref.~\cite{Cai2013}. It is possible that hidden logarithmic corrections could also have been overlooked in these studies, and that the problem deserves further attention.

For a more general setting such as a Lindbladian system with an interacting integrable Hamiltonian, the associated t-GGE ensemble is generally non-Gaussian, which makes the problem challenging to treat analytically. Nevertheless, the t-GGE approach has successfully been applied to a Lieb-Liniger gas with finite interaction strength and local $K$-body losses (with $K=1,2,3$)~\cite{Bouchoule2020}. In principle, a similar framework could be used to study a non-reciprocal open quantum system with coherent dynamics described by the XXZ model, although a simple set of closed equations such as those in Eq.~\eqref{eq:GGEeq} are not expected to exist.

More generally, this work suggests the importance of considering with appropriate theoretical tools that go beyond simple mean-field treatments the dynamics of integrable models, such as the XX quantum spin chain, when integrability is weakly broken, in this case by non-reciprocal losses. This poses the intriguing question of how to study the action of non-reciprocal losses in the case of a quantum spin chain that does not support the Hamiltonian magnetization current as a conservation law.
Such models exist, and many non-integrable models have this property.
Whether external reservoirs could induce a current flow also in those situations is an intriguing question that deserves to be studied.

%%%%%%%%%%%%%%%%%%%%%%%%%%%%%%%%%%%%%%%%%%%%%%%%%%%%%%%%%%%%%%%%%%%%%%%%%%%%%%%%%%%%%%%%%%%%%%%%%%%%%%%%%%%%%%%
%%%%%%%%%%%%%%%%%%%%%%%%%%%%%%%%%%%%%%%%%%%%%%%%%%%%%%%%%%%%%%%%%%%%%%%%%%%%%%%%%%%%%%%%%%%%%%%%%%%%%%%%%%%%%%%
\section*{Acknowledgements}
%%%%%%%%%%%%%%%%%%%%%%%%%%%%%%%%%%%%%%%%%%%%%%%%%%%%%%%%%%%%%%%%%%%%%%%%%%%%%%%%%%%%%%%%%%%%%%%%%%%%%%%%%%%%%%%
%%%%%%%%%%%%%%%%%%%%%%%%%%%%%%%%%%%%%%%%%%%%%%%%%%%%%%%%%%%%%%%%%%%%%%%%%%%%%%%%%%%%%%%%%%%%%%%%%%%%%%%%%%%%%%%

We thank  G.~Aupetit-Diallo, S.~E.~Begg, M.~Brunelli, J.~Dubail, R.~Hanai, L.~Lumia, and M.~Schirò for enlightening discussions. We are extremely grateful to S.~E.~Begg and R.~Hanai for sharing their numerical data.
AM, AN and LM thank the University of Tokyo for warm hospitality.
We thank R.~Daraban, A.~Safavi-Naini, J.~Schachenmayer and M.~Maghrebi for enlightening discussions on the subject and on their work, Ref.~\cite{daraban2026}.

\paragraph{Funding information}

This work was carried out in the framework of the joint Ph.D.\ program between the CNRS and the University of Tokyo. This work is supported by the ANR project LOQUST ANR-23-CE47-0006-02 and by the PEPR Dyn-1D ANR-23-PETQ-0001.
This work is part of HQI (www.hqi.fr) initiative and is supported by France 2030 under the French National Research Agency grant number ANR-22-PNCQ-0002.
H.K. was supported by JSPS KAKENHI Grants No. JP23K25783, No. JP23K25790, and MEXT KAKENHI Grant-in-Aid for Transformative Research Areas A “Extreme Universe” (KAKENHI Grant No. JP21H05191).

\begin{appendix}

\section{Proof that $\langle c^\dag(k) c(q) \rangle =0$ for $k\neq q$} \label{ap:TI}
In this appendix, we show that $\langle c^\dag(k) c(q) \rangle =0$, for $k\neq q$, is a direct consequence of the translational invariance of the system.
By going back to real space, we find
\begin{equation}
  2 \pi \langle c^\dag(k) c(q) \rangle  = \sum_{j=-\infty}^{\infty} \sum_{l=-\infty}^{\infty} e^{ikj} e^{-iql} \langle c_j^\dag c_l \rangle
  =  \sum_{j=-\infty}^\infty  \sum_{\eta=-\infty}^{\infty} e^{i(k-q)j} e^{-iq\eta} \langle c_j^\dag c_{j+\eta} \rangle
\end{equation}
From the translational invariance, we obtain that, for all $\eta \in \mathbb{Z}$, $\langle c_j^\dag c_{j + \eta} \rangle =: f(\eta)$ is independent of the lattice site. Thus,
\begin{equation}
  2 \pi \langle c^\dag(k) c(q) \rangle = \underbrace{\left[ \sum_{j=-\infty}^\infty e^{i(k-q)j} \right]}_{=0 \text{ if } k \neq q} \left[ \sum_{\eta=-\infty}^{\infty} f(\eta)  e^{iq\eta}  \right]
\end{equation}
Therefore, if  $k\neq q$ then $\langle c^\dag(k) c(q) \rangle =0$. This justifies the form of the Eq.~\eqref{eq:defrho}.

%%%%%%%%%%%%%%%%%%%%%%%%%%%%%%%%%%%%%%%%%%%%%%%%%%%%%%%%%%%%%%%%%%%%%%%%%%%%%%%%%%%%%%%%%%%%%%%%%%%%%%%%%%%%%%%
%%%%%%%%%%%%%%%%%%%%%%%%%%%%%%%%%%%%%%%%%%%%%%%%%%%%%%%%%%%%%%%%%%%%%%%%%%%%%%%%%%%%%%%%%%%%%%%%%%%%%%%%%%%%%%%

\section{Derivation of Eq.~\eqref{eq:IC}} \label{ap:rho0}

In this appendix, we show how to obtain the momentum distribution associated with the initial state~\eqref{eq:Psi0}, written as \eqref{Eq:Initial:State} in the fermionic language. 
First we note that
\begin{equation}
    \underbrace{\delta(k-q)}_{\frac{1}{2 \pi } \sum_j e^{i(k-q)j}} \varrho_0(k) =  \bra{\Psi_0} c^\dag(k) c(q) \ket{\Psi_0}  = \frac{1}{2 \pi } \sum_{j,l=-\infty}^{+\infty} e^{ikj} e^{-iql} \bra{\Psi_0} c_j^\dag c_l \ket{\Psi_0} .\label{eq:corr_rho0}
\end{equation} 
From the fermionic expression of the initial state $\ket{\Psi_0}$, we can show that
\begin{equation}
\begin{split}
        \bra{\Psi_0} c_j^\dag c_l \ket{\Psi_0} &= \bra{\rm vac} \ldots \Big(\cos \theta c_2 + \sin \theta \Big)  \Big(\cos \theta c_1 + \sin \theta \Big) \ldots c_j^\dag c_l   \\ & \qquad \quad \  \ldots\left(\cos \theta c_1^\dag + \sin \theta \right)  \left(\cos \theta c_2^\dag + \sin \theta \right) \ldots \ket{\rm vac} \\&= \begin{cases}
        \cos^2(\theta), &\text{ if } j=l,\\
        \cos^2(\theta) \sin^2(\theta) \left[ \sin^2(\theta) -\cos^2(\theta)\right]^{|j-l|-1}, & \text{ if } j\neq l.
    \end{cases} \label{eq:corr_Psi0}
\end{split}
\end{equation}
Using Eqs.~\eqref{eq:corr_rho0} and \eqref{eq:corr_Psi0}, we obtain Eq.~\eqref{eq:IC} after a few algebraic manipulations.

\section{Derivation of Eq.~\eqref{eq:chi_ff}} \label{ap:freefermions}

In this appendix, we provide an analytical proof of the late-time algebraic decay $n \sim t^{-\chi}$ for the free fermionic system, where the critical exponent $\chi$ is given in Eq.~\eqref{eq:chi_ff}. 

From the Eq.~\eqref{eq:n_ff} and by doing the change of variable $\varepsilon = k - \kbar$ with $\kbar = \pi-\phi$, we find
\begin{equation}
\begin{split}
n &= \int_{-\kbar}^{2 \pi-\kbar} \frac{d \varepsilon}{ 2 \pi} \varrho_0(\varepsilon+ \kbar) e^{- 2 \left( 1-\cos(\varepsilon) \right) \kappa t} \\ & \underset{\kappa t \gg 1}{\simeq} \int_{-\infty}^{+\infty} \frac{d \varepsilon}{2\pi} \left( \varrho_0(\kbar)  +  \varepsilon \frac{\partial \varrho_0}{\partial k}\Bigr|_{\substack{k=\kbar}} +  \frac{\varepsilon^2}{2} \frac{\partial^2 \varrho_0}{\partial k^2}\Bigr|_{\substack{k=\kbar}}   \right) e^{- \varepsilon^2 \kappa t}.
\end{split}
\end{equation}
In the second line, we have performed a saddle point approximation around $\varepsilon = 0$.
We then remark that,
\begin{equation}
\begin{split}
          \varrho_0(\kbar) \neq 0 &\text{ if } \phi \neq 0 \text{ or } \theta=0 \ \text{ (case 1)},\\   \varrho_0(\kbar) = \frac{\partial \varrho_0}{\partial k}\Bigr|_{\substack{k=\kbar}}= 0, \ \frac{\partial^2 \varrho_0}{\partial k^2}\Bigr|_{\substack{k=\kbar}} \neq 0 & \text{ if }  \phi = 0 \text{ and } \theta \in \left( 0, \frac{\pi}{2}\right) \ \text{ (case 2)}.
\end{split}
\end{equation}
In the case 1, we consider only the dominant term of order $O(\varepsilon^0)$ in the Taylor expansion of $\varrho_0(\varepsilon+ \kbar)$, thus
\begin{equation}
    n \simeq  \varrho_0(\kbar) \int_{-\infty}^{+\infty} \frac{d \varepsilon}{2\pi}  e^{- \varepsilon^2 \kappa t} = \varrho_0(\kbar) \frac{1}{2 \sqrt{ \pi \kappa t} }\propto t^{-1/2}.
\end{equation}
In the case 2, the dominant term in the Taylor expansion of $\varrho_0(\varepsilon+ \kbar)$ is of the order $O(\varepsilon^2)$, thus
\begin{equation}
    n \simeq \frac{\partial^2 \varrho_0}{\partial k^2}\Bigr|_{\substack{k=\kbar}} \int_{-\infty}^{+\infty} \frac{d \varepsilon}{2\pi} \frac{\varepsilon^2}{2} e^{- \varepsilon^2 \kappa t} = \frac{\partial^2 \varrho_0}{\partial k^2}\Bigr|_{\substack{k=\kbar}} \frac{1}{8 \sqrt{\pi}\left(  \kappa t\right)^{3/2} }\propto t^{-3/2}.
\end{equation}

Finally, we mention that $n$ can be written compactly in terms of the modified Bessel functions of the first kind $I_\alpha(x)$, in the cases $\theta =0,\frac{\pi}{4}$. 
By using the substitution
\begin{equation}
    e^{-2\kappa \cos(\phi+k) t} = \sum_{\alpha=-\infty}^{\infty} I_\alpha(-2\kappa t) e^{i \alpha (\phi+k)}
\end{equation}
in Eq.~\eqref{eq:n_ff}, we can show that
\begin{equation}
    n = \begin{cases}
    e^{-2 \kappa t} I_0(2 \kappa t), & \text{ if } \theta=0,\\ \\
    \dfrac{e^{-2\kappa t}}{2} \left[ I_0(2 \kappa t) - \cos(\phi) I_1(2\kappa t)\right],&\text{ if } \theta = \dfrac{\pi}{4}.
    \end{cases}
\end{equation}
The latter expressions are valid at any times. The asymptotic expansion of the modified Bessel functions reads 
\begin{equation}
    I_\alpha(x) = \dfrac{e^x}{\sqrt{2  \pi x }} \left(  1-\dfrac{4 \alpha^2-1}{8x} + O\left( x^{-2}\right)\right).
\end{equation}
Thus, for $\kappa t \gg 1$, we obtain,
\begin{equation}
    n \sim \begin{cases}
    \dfrac{1}{2 \sqrt{\pi \kappa t}} \left( 1+ \dfrac{1}{16 \kappa t}\right), & \text{ if } \theta=0,\\ \\
    \dfrac{1}{4 \sqrt{\pi \kappa t}} \left( 1-\cos \phi +\dfrac{1+ 3 \cos \phi}{16 \kappa t}\right),&\text{ if } \theta = \dfrac{\pi}{4}.
    \end{cases}
\end{equation}
By keeping the first non-vanishing order in $(\kappa t)^{-1}$, we find again consistent results with Eq.~\eqref{eq:chi_ff}.
\section{Derivation of Eq.~\eqref{eq:GGEeq}} \label{ap:Ap1}

In this appendix, we derive Eq.~\eqref{eq:GGEeq}.
We start from Eq.~\eqref{eq:evolcc2}, and we compute independently $\langle L_{\jj}^\dag L_{\jj} c^\dag (k) c(k) \rangle$ and $\langle L_{\jj}^\dag c^\dag (k) c(k)  L_{\jj} \rangle$; 
the expression of the operator~$L_{\jj}$ is given in Eq.~\eqref{eq:jumps_fermions}.

First, we focus on $\langle L_{\jj}^\dag L_{\jj} c^\dag (k) c(k) \rangle$. For this term, the presence of a string operator in $L_{\jj}$'s expression does not play any role because
\begin{equation}
 \begin{split}
L_{\jj}^\dag L_{\jj} & = \left( c_{\jj}^\dag + e^{- i \phi} c_{{\jj}+1}^\dag (-1)^{n_{\jj}} \right)  \left( c_{\jj} + e^{i \phi}   (-1)^{n_{\jj}} c_{{\jj}+1} \right)\\ & = c_{\jj}^\dag c_{\jj} + e^{i \phi} c_{{\jj}}^\dag (-1)^{n_{\jj}} c_{{\jj}+1} + e^{-i \phi} c_{{\jj}+1}^\dag (-1)^{n_{\jj}} c_{\jj}+ c_{{\jj}+1}^\dag c_{{\jj}+1} \\ & =  c_{\jj}^\dag c_{\jj} + e^{i \phi} c_{{\jj}}^\dag  c_{{\jj}+1} + e^{-i \phi} c_{{\jj}+1}^\dag c_{\jj} + c_{{\jj}+1}^\dag c_{{\jj}+1}.
\end{split}
\end{equation}
By going in real space, we obtain
\begin{equation}
 \begin{split}
\langle L_{\jj}^\dag L_{\jj} c^\dag(k) c(k) \rangle & = \frac{1}{2 \pi} \sum_{n,l=-\infty}^{+\infty} e^{ik(l-n)} \langle L_{\jj}^\dag L_{\jj} c_l^\dag c_n \rangle \\ & = \frac{1}{2 \pi} \sum_{n,l=-\infty}^{+\infty} e^{ik(l-n)} \Big( \langle c_{\jj}^\dag c_{\jj}  c_l^\dag c_n \rangle + e^{i \phi} \langle c_{{\jj}}^\dag  c_{{\jj}+1}   c_l^\dag c_n \rangle  \\ & \qquad \quad \qquad  \quad \qquad + e^{-i \phi} \langle c_{{\jj}+1}^\dag c_{\jj}  c_l^\dag c_n \rangle   +  \langle c_{{\jj}+1}^\dag c_{{\jj}+1}  c_l^\dag c_n \rangle \Big). \label{eq:LL}
\end{split}
\end{equation}
Using Eq.~\eqref{eq:LL} and the following identities
\begin{subequations}
 \begin{equation}
 \langle c_{\jj}^\dag c_m c^\dag_l c_n \rangle = \langle c_{\jj}^\dag c_m \rangle \langle c^\dag_l c_n \rangle + \langle c_{\jj}^\dag c_n \rangle \langle c_m c^\dag_l \rangle \qquad \text{ (Wick theorem), }
\end{equation}
\begin{equation}
  \langle c_{\jj}^\dag c_m \rangle = \frac{1}{2 \pi}\int_0^{2 \pi} dq \int_0^{2 \pi} dp e^{-iq {\jj}}e^{ipm} \underbrace{\langle c^\dag(q) c(p)\rangle}_{\delta(q-p) \varrho(q)} = \int_0^{2 \pi} \frac{dq}{2 \pi} e^{iq(m-{\jj})} \varrho(q),\label{eq:3b}
\end{equation}
\begin{equation}
  \langle c_m c_{\jj}^\dag \rangle = \frac{1}{2 \pi}\int_0^{2 \pi} dq \int_0^{2 \pi} dp e^{-iq {\jj}}e^{ipm} \underbrace{ \langle c(p)  c^\dag(q) \rangle}_{\delta(q-p) (1-\varrho(q))} = \int_0^{2 \pi} \frac{dq}{2 \pi} e^{iq(m- {\jj})} (1-\varrho(q)),\label{eq:3c}
\end{equation}
\begin{equation}
 \sum_{l=-\infty}^{\infty} e^{i K l} = 2 \pi \sum_{m = -\infty}^{\infty} \delta (K - 2 \pi m),
\end{equation}
\end{subequations}
we find
\begin{equation}
 \begin{split}
 \pi \langle L_{\jj}^\dag L_{\jj} c^\dag(k) c(k) \rangle
&=  \int_0^{2 \pi} \frac{dq}{2 \pi}  \varrho(q) \left(1 + 1 \cos(q+\phi) \right) \int_0^{2 \pi} \frac{dp}{2 \pi} \varrho(p) \left[ \sum_{l} e^{i(k-p)l}\right] \left[ \sum_{n} e^{i(p-k)n}\right] \\ &\quad +  \varrho(k) (1-\varrho(k)) (1+\cos(k+\phi)).
\end{split}
\end{equation}
The first term on the right-hand side of the latter expression diverges, however this divergence disappears when we subtract the contribution $\pi \langle L_{\jj}^\dag c^\dag(k) c(k)  L_{\jj}  \rangle$. Note that the Wick theorem could be applied here, because we assume that the system's density matrix is well approximated by the t-GGE~\eqref{eq:GGEstate} which is gaussian.

We now focus on the other term $\langle L_{\jj}^\dag c^\dag(k) c(k) L_{\jj}\rangle$ for which we have to take care of the string operators $(-1)^{n_{\jj}}$ and $(-1)^{N_{(-\infty, {\jj}-1)}}$.
We have
\begin{equation}
 \begin{split}
&\langle L_{\jj}^\dag  c^\dag(k) c(k) L_{\jj} \rangle = \\& = \langle \left( c_{\jj}^\dag + e^{- i \phi} c_{{\jj}+1}^\dag (-1)^{n_{\jj}} \right) (-1)^{N_{(-\infty, {\jj}-1)}}  c^\dag(k) c(k)  (-1)^{N_{(-\infty, {\jj}-1)}}  \left( c_{\jj} + e^{ i \phi} (-1)^{n_{\jj}} c_{{\jj}+1}  \right) \rangle \\ &= \frac{1}{2 \pi} \sum_{l,n = -\infty}^\infty e^{ik(l-n)}  \langle \left( c_{\jj}^\dag + e^{- i \phi} c_{{\jj}+1}^\dag (-1)^{n_{\jj}} \right)      (-1)^{N_{(-\infty, {\jj}-1)}}  c^\dag_l c_n  \\ & \hspace{7cm} \times (-1)^{N_{(-\infty, {\jj}-1)}}  \Bigl( c_{\jj} + e^{ i \phi} (-1)^{n_{\jj}} c_{{\jj}+1}  \Bigr) \rangle.
\end{split} \label{eq:eq1}
\end{equation}
The string operator induces sign functions, which we denote $s(l-{\jj})$ and $\mu_{{\jj},l}$:
\begin{subequations}
\begin{equation}
 (-1)^{N_{(-\infty, {\jj}-1)}} c_l^\dag = \sg(l-{\jj}) c_l^\dag (-1)^{N_{(-\infty, {\jj}-1)}}, \quad \text{ with } \quad \sg(l-{\jj}) =
\begin{cases}
 \ 1 \text{  if } l \geq {\jj},\\
-1 \text{ if } l < {\jj},
\end{cases} \label{eq:sgn}
\end{equation}
\begin{equation}
 (-1)^{n_{\jj}} c_l^\dag = \mu_{{\jj},l} c_{l}^\dag (-1)^{n_{\jj}}, \quad  c_l (-1)^{n_{\jj}} = \mu_{{\jj},l} (-1)^{n_{\jj}} c_{l}, \quad \text{ with } \quad \mu_{{\jj},l} =
\begin{cases}
 \ 1 \text{  if } l \neq {\jj},\\
-1 \text{ if } l = {\jj}.
\end{cases}
\end{equation} \label{eq:6}
\end{subequations}
Using Eqs.~\eqref{eq:eq1}, \eqref{eq:6}, and that $\mu_{l,{\jj}}\sg(l-{\jj}) = -\sg({\jj}-l)$, we obtain
\begin{equation}
 \begin{split}
\langle L_{\jj}^\dag  c^\dag(k) c(k) L_{\jj} \rangle & = \frac{1}{2 \pi} \sum_{l,n}  e^{ik(l-n)} \sg(l-{\jj}) \sg(n-{\jj})  \left( \langle c_{\jj}^\dag c_l^\dag c_n c_{\jj} \rangle  + \langle c_{{\jj}+1}^\dag c_l^\dag c_n c_{{\jj}+1} \rangle \right)  \\ &   \quad +\frac{1}{2 \pi} \sum_{l,n} e^{ik(l-n)} \sg({\jj}-l) \sg({\jj}-n) \left( \langle c_{\jj}^\dag c_l^\dag c_n c_{{\jj}+1} \rangle  e^{i \phi} + \langle c_{{\jj}+1}^\dag  c_l^\dag c_n c_{{\jj}} \rangle  e^{-i \phi} \right). \label{eq:eq6}
\end{split}
\end{equation}
From the Eqs.~\eqref{eq:eq6}, \eqref{eq:3b}, \eqref{eq:3c}, the Wick theorem, and  the identity
\begin{equation}
 \forall K, \quad \sum_{l=-\infty}^{\infty} \sg(l-{\jj}) e^{iKl}  = \frac{2 e^{iK{\jj}}}{1-e^{iK}}= \frac{i e^{iK{\jj}} e^{-iK/2}}{\sin(K/2)} = e^{iK{\jj}} \left( 1+ i \cot(K/2) \right),
\end{equation}
we obtain
\begin{subequations}
 \begin{equation}
 \begin{split}
\pi  \langle L_{\jj}^\dag  c^\dag(k) c(k) L_{\jj} \rangle = A(k) \ + &\int_0^{2 \pi} \frac{dq}{2 \pi}  \varrho(q) \left(1 +  \cos(q+\phi) \right)  \\ &  \times \int_0^{2 \pi}\frac{dp}{2 \pi} \varrho(p) \left[ \sum_{l} \sg(l-{\jj}) e^{i(k-p)l}\right] \left[ \sg(n-{\jj}) \sum_{n} e^{i(p-k)n}\right],
 \end{split}
\end{equation}
 \begin{equation}
  \begin{split}
\text{with} \quad A(k) := & -   \int  \frac{dq}{2 \pi} \varrho(q)  \int  \frac{dp}{2 \pi} \varrho(p) \left(1+\cos(p+\phi) \right) \\
&  - \fint  \frac{dq}{2 \pi} \varrho(q)  \cot \left(\frac{k-q}{2} \right)  \fint  \frac{dp}{2 \pi} \varrho(p) \cot \left(\frac{k-p}{2}\right)  \left( 1+ \cos(p+\phi) \right) \\
& +  \fint  \frac{dq}{2 \pi} \varrho(q)   \fint  \frac{dp}{2 \pi} \varrho(p) \sin(p+\phi)  \left[\cot \left(\frac{k-q}{2}\right) - \cot \left(\frac{k-p}{2}\right)   \right] \\
=& - 2 \left(  \fint_{0}^{2 \pi} \frac{dq}{2 \pi} \varrho(q) \frac{\cos \left( \frac{q+\phi}{2}\right)}{\sin \left( \frac{k-q}{2}\right)}\right)^2. \label{eq:12b}
\end{split}
\end{equation}
\end{subequations}
Therefore,
\begin{subequations}
 \begin{equation}
\begin{split}
&\pi  \left[ \langle L_{\jj}^\dag L_{\jj} c^\dag(k) c(k) \rangle - \langle L_{\jj}^\dag c^\dag(k) c(k) L_{\jj} \rangle \right]   \\ &=   \varrho(k) \left( 1- \varrho(k) \right) \left( 1 +  \cos(k + \phi) \right)  - A(k) \ +  \int \frac{dq}{2 \pi} \varrho(q)  \left(1+\cos(q+\phi) \right) \fint  \frac{dp}{2 \pi} \varrho(p) Y(k-p), \label{eq:13a}
\end{split}
\end{equation}
\begin{equation}
     \text{ with } \quad Y(K) := \left| \sum_{l=-\infty}^{\infty} e^{iKl}  \right|^2 - \left|\sum_{l=-\infty}^{\infty} s(l) e^{iKl} \right|^2 = \sum_{m=-\infty}^\infty e^{i K m}  \underbrace{\sum_{n=-\infty}^\infty (1-s(m+n)s(n))}_{=2|m|}.
\end{equation}
\end{subequations}
By decomposing $\varrho(p)$ in Fourier modes, $\varrho(p) = \sum_{m} \varrho_m e^{imp}$, we can show that
\begin{equation}
 \fint_0^{2 \pi} dp \varrho(p) Y(k-p)= \fint_0^{2 \pi} dp \frac{\varrho(k) - \varrho(p)}{\sin^2\left( \frac{k-p}{2}\right)}.
\end{equation}
The latter expression, together with Eqs.~\eqref{eq:12b},~\eqref{eq:13a} and \eqref{eq:evolcc2}, leads to Eq.~\eqref{eq:GGEeq}, or alternatively Eq.~\eqref{eq:GGEeqNum}.

%%%%%%%%%%%%%%%%%%%%%%%%%%%%%%%%%%%%%%%%%%%%%%%%%%%%%%%%%%%%%%%%%%%%%%%%%%%%%%%%%%%%%%%%%%%%%%%%%%%%%%%%%%%%%%%
%%%%%%%%%%%%%%%%%%%%%%%%%%%%%%%%%%%%%%%%%%%%%%%%%%%%%%%%%%%%%%%%%%%%%%%%%%%%%%%%%%%%%%%%%%%%%%%%%%%%%%%%%%%%%%%
\section{Details on the numerical solution of Eq.~\eqref{eq:GGEeqNum}} \label{ap:Ap2}

To solve numerically Eq.~\eqref{eq:GGEeqNum}, we discretize the momentum interval $[0,2\pi]$ and use Runge–Kutta method. We also use the following relations between the circular Hilbert transform and the Fourier transform. By defining the Fourier transform and its inverse as
\begin{equation}
 \mathcal{F}[f](n) =  \int_0^{2 \pi} \frac{dx}{2 \pi} f(x) e^{- i n x}  := \hat{f}_n, \qquad \mathcal{F}^{-1}[g](x) =  \sum_{n=-\infty}^{+ \infty}  g_n e^{inx},
\end{equation}
we have
\begin{equation}
 \mathcal{F} [ \mathcal{H}[f]](n) = -i \text{sgn}(n) \hat{f}_n \quad \text{ where } \quad \text{sgn}(n) =  \begin{cases}
 \ 1 \text{  if } n>0\\
  0 \text{  if } n=0\\
-1 \text{ if } n<0,
\end{cases}  \label{eq:FourierHilbert1}
\end{equation}
\begin{equation}
 \mathcal{F} [ \mathcal{H}[f]^\prime](n) = |n| \hat{f}_n. \label{eq:FourierHilbert2}
\end{equation}
Inverting Eqs.~\eqref{eq:FourierHilbert1} and \eqref{eq:FourierHilbert2} provides an efficient method to compute a circular Hilbert transform or its derivative using fast Fourier transform.

\begin{figure}[t]
\centering
 \includegraphics[width=\columnwidth]{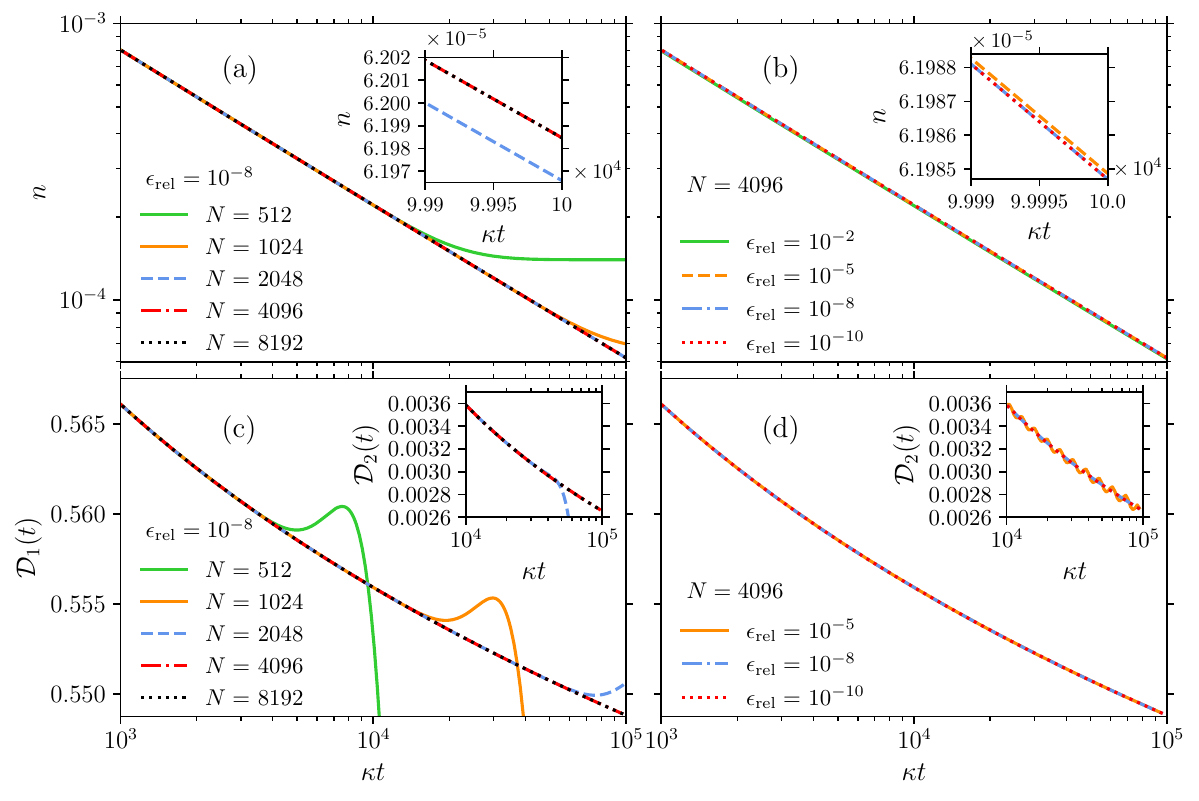}
 \caption{(a--b) Time evolution of the magnetization $n$ for various values of $N$ and $\epsilon_{\rm rel}$; the insets show the convergence at very large times. (c--d) Time evolution of the first logarithmic derivative $\mathcal{D}_1$ for various values of $N$ and $\epsilon_{\rm rel}$; the insets show the time evolution of the second logarithmic derivative $\mathcal{D}_2$ . For all panels, we use $\phi = - \pi/2$, $\theta=0$, and $\epsilon_{\rm abs} = 10^{-4} \epsilon_{\rm rel}$.}
\label{fig:numrical_stability}
\end{figure}

In practice, we use the function \texttt{solve\_ivp} from the Python library SciPy, which uses an adaptive time step. We assess the convergence of the numerical solution by varying three parameters: $N$, the number of equally spaced discretization points in the momentum interval  $[0,2\pi]$, and $\epsilon_{\rm rel}$ and $\epsilon_{\rm abs}$, the relative and absolute error tolerances, respectively. We choose $N \in \{ 512 , 1024, 2048,4096,8192\}$, $\epsilon_{\rm rel} \in \{ 10^{-2}, 10^{-5}, 10^{-8}, 10^{-10}\}$, and $\epsilon_{\rm abs} = 10^{-4} \epsilon_{\rm rel}$. In Fig.~\ref{fig:numrical_stability}, we show the time evolutions of both the magnetization $n$ and its first and second logarithmic derivatives $\mathcal{D}_1$ and $\mathcal{D}_2$, for $\theta=0$ and $\phi=-\pi/2$. We have checked that, for all the values of $\theta$ and $\phi$ considered in this article, the numerical solution has converged. In Figs.~\ref{fig:n}, \ref{fig:log_der} and \ref{fig:dn}, displayed in the main text, we choose $N=4096$ and $\epsilon_{\rm rel}=10^{-8}$.

%%%%%%%------FIG---------%%%%%%%
%%%%%%%-----------------%%%%%%%

%%%%%%%%%%%%%%%%%%%%%%%%%%%%%%%%%%%%%%%%%%%%%%%%%%%%%%%%%%%%%%%%%%%%%%%%%%%%%%%%%%%%%%%%%%%%%%%%%%%%%%%%%%%%%%%
%%%%%%%%%%%%%%%%%%%%%%%%%%%%%%%%%%%%%%%%%%%%%%%%%%%%%%%%%%%%%%%%%%%%%%%%%%%%%%%%%%%%%%%%%%%%%%%%%%%%%%%%%%%%%%%
\section{Details on the finite size numerical computation via quantum trajectories} \label{ap:Ap3}
%%%%%%%%%%%%%%%%%%%%%%%%%%%%%%%%%%%%%%%%%%%%%%%%%%%%%%%%%%%%%%%%%%%%%%%%%%%%%%%%%%%%%%%%%%%%%%%%%%%%%%%%%%%%%%%
%%%%%%%%%%%%%%%%%%%%%%%%%%%%%%%%%%%%%%%%%%%%%%%%%%%%%%%%%%%%%%%%%%%%%%%%%%%%%%%%%%%%%%%%%%%%%%%%%%%%%%%%%%%%%%%

In this appendix, we provide details about how to deal with a finite system size, with periodic boundary conditions (PBC), as it is the case in our quantum trajectory simulations. This is also the setup considered in Refs.~\cite{Bouchoule2020,Riggio2024,Lumia2025}.

For a finite system with lattice site $j \in \{ 1, 2, \ldots, L\}$, the Jordan-Wigner transform is defined as
\begin{equation}
 c_j =  (-1)^{N_{(1,j-1)}} S_j^-, \qquad c_j^\dag = (-1)^{N_{(1,j-1)}} S_j^+, \qquad \text{ with } \quad N_{(1,j)} = \sum_{l = 1}^{j} n_l. \label{eq:JW_2}
\end{equation}
Importantly, we assume PBC for the spin operators, \textit{i.e. $S_{L+1}^+ = S_1^+$}; this implies that the fermionic operators satisfy the boundary condition $c_{L+1}^\dag = (-1)^{N_{(1,L)} -1} c_1^\dag $. As a consequence, when going in Fourier space two different sets of momenta should be considered:
\begin{equation}
   \Qp = \frac{2 \pi}{L} \{ 1,2,\ldots, L \}, \quad \Qap = \frac{2 \pi}{L} \left\{ \frac{1}{2},\frac{3}{2},\ldots, L - \frac{1}{2} \right\}.
\end{equation}
Intuitively, the fermions satisfy PBC in the sector where the total number of particles is odd and anti PBC in the sector where the  number of particles is even. In practice, the Fourier modes are expressed as
\begin{equation}
    c_j^\dag = \frac{1}{\sqrt{L}} \sum_{k \in \Qap} e^{-ikj}c^\dag(k) = \frac{1}{\sqrt{L}} \sum_{k \in \Qp} e^{-ikj}c^\dag(k), \quad  \ \forall k \in Q, \ c^\dag(k) = \frac{1}{\sqrt{L}} \sum_{j=1}^L e^{ikj}c_j^\dag,
\end{equation}
where $Q := \Qap \cup \Qp$, and the Hamiltonian is
\begin{equation}
    H = - \frac{J}{2} \sum_{j=1}^L \left( S_{j+1}^+ S_j^- + S_j^+ S_{j+1}^- \right) = \sum_{k \in \Qap} \epsilon_k P_+ c^\dag(k) c(k) + \sum_{k \in \Qp} \epsilon_k P_- c^\dag(k) c(k),
\end{equation}
with
\begin{equation}
   \epsilon_k = -J\cos k, \quad \text{ and } \quad P_\pm = \frac{1}{2} \left( 1 \pm (-1)^{N_{(1,L)}} \right).
\end{equation}
The finite system size momentum distribution is defined as
\begin{equation}
 \forall k \in \Qap, \ \varrho_{\rm ap} (k)  = \langle P_+ c^\dag(k) c(k) \rangle, \quad \forall k \in \Qp, \ \varrho_{\rm p} (k)  = \langle P_{-} c^\dag(k) c(k) \rangle,
\end{equation}
\begin{equation}
   \forall n \in \{1, \ldots, L \}, \  \tilde{\varrho} \big( 2 \pi (n-1/4)/L \big)  =  \varrho_{\rm ap} \big( 2 \pi (n-1/2)/L \big) + \varrho_{\rm p} \big( 2 \pi n/L \big). 
\end{equation}
The crosses in Fig.~\ref{fig:rhok} correspond to the distribution $\tilde{\varrho}$, computed using quantum trajectories for $L=14$. Empirically, we found that $\tilde{\varrho}$ is a good proxy for the comparison with the t-GGE results.

%%%%%%%%%%%%%%%%%%%%%%%%%%%%%%%%%%%%%%%%%%%%%%%%%%%%%%%%%%%%%%%%%%%%%%%%%%%%%%%%%%%%%%%%%%%%%%%%%%%%%%%%%%%%%%%
%%%%%%%%%%%%%%%%%%%%%%%%%%%%%%%%%%%%%%%%%%%%%%%%%%%%%%%%%%%%%%%%%%%%%%%%%%%%%%%%%%%%%%%%%%%%%%%%%%%%%%%%%%%%%%%
\section{Details on the profile of $w_r$} 
%%%%%%%%%%%%%%%%%%%%%%%%%%%%%%%%%%%%%%%%%%%%%%%%%%%%%%%%%%%%%%%%%%%%%%%%%%%%%%%%%%%%%%%%%%%%%%%%%%%%%%%%%%%%%%%
%%%%%%%%%%%%%%%%%%%%%%%%%%%%%%%%%%%%%%%%%%%%%%%%%%%%%%%%%%%%%%%%%%%%%%%%%%%%%%%%%%%%%%%%%%%%%%%%%%%%%%%%%%%%%%%

In this appendix, we provide more analytical details on the profile of $w_r(t)$ as defined in~Eq.~\eqref{eq:def_w_r}.

%%%%%%%%%%%%%%%%%%%%%%%%%%%%%%%%%%%%%%%%%%%%%%%%%%%%%%%%%%%%%%%%%%%%%%%%%%%%%%%%%%%%%%%%%%%%%%%%%%%%%%%%%%%%%%%
%%%%%%%%%%%%%%%%%%%%%%%%%%%%%%%%%%%%%%%%%%%%%%%%%%%%%%%%%%%%%%%%%%%%%%%%%%%%%%%%%%%%%%%%%%%%%%%%%%%%%%%%%%%%%%%
\subsection{Free fermions} \label{ap:w_r_ff}
%%%%%%%%%%%%%%%%%%%%%%%%%%%%%%%%%%%%%%%%%%%%%%%%%%%%%%%%%%%%%%%%%%%%%%%%%%%%%%%%%%%%%%%%%%%%%%%%%%%%%%%%%%%%%%%
%%%%%%%%%%%%%%%%%%%%%%%%%%%%%%%%%%%%%%%%%%%%%%%%%%%%%%%%%%%%%%%%%%%%%%%%%%%%%%%%%%%%%%%%%%%%%%%%%%%%%%%%%%%%%%%

In the case of the free fermions, we can perform a computation in the same spirit as in App.~\ref{ap:freefermions}. We directly substitute the solution of Eq.~\eqref{eq:rhok_ff}
\begin{equation}
    \vr(k,t) = \vr_0(k) e^{- 2  \left( 1 - \cos(k-\bar{k}) \right) \kappa t}
\end{equation}
into Eq.~\eqref{eq:def_w_r}, to obtain
\begin{equation}
\begin{split}
    w_r(t) &= \int_0^{2 \pi} \frac{dk}{2 \pi} \vr_0(k) e^{- 2  \left( 1 - \cos(k-\bar{k}) \right) \kappa t} \cos((k-\bar{k})r) \\ & \underset{\kappa t \gg 1}{\simeq} \int_{-\infty}^{+\infty} \frac{d \varepsilon}{2 \pi}  \left( \varrho_0(\kbar)  +  \varepsilon \frac{\partial \varrho_0}{\partial k}\Bigr|_{\substack{k=\kbar}} +  \frac{\varepsilon^2}{2} \frac{\partial^2 \varrho_0}{\partial k^2}\Bigr|_{\substack{k=\kbar}}   \right) e^{- \varepsilon^2 \kappa t}   \cos(\varepsilon r) 
    \\  & \underset{\kappa t \gg 1}{\simeq} \frac{1}{2\sqrt{ \pi \kappa t}}e^{-\frac{r^2}{4 \kappa t}} \left[ \varrho_0(\kbar) +  \frac{1}{4 \kappa t}  \left( 1 - \frac{r^2}{2 \kappa t}\right) \frac{\partial^2 \varrho_0}{\partial k^2}\Bigr|_{\substack{k=\kbar}} \right]
\end{split}
\end{equation}
At the second line, we have used the change of variable $\varepsilon = k - \bar{k}$, and a saddle point approximation around the point $\varepsilon=0$. 
Moreover, $ \varrho_0(\kbar) =0 $ and $\frac{\partial^2 \varrho_0}{\partial k^2}\Bigr|_{\substack{k=\kbar}} \neq 0$, if $\phi = 0$ and $\theta \in \left( 0, \frac{\pi}{2}\right)$, while $\varrho_0(\kbar) \neq 0$ otherwise. Thus, by keeping the first non vanishing order in $(\kappa t)^{-1}$, we obtain Eq.~\eqref{eq:w_r_gaussian_ff}.

%%%%%%%%%%%%%%%%%%%%%%%%%%%%%%%%%%%%%%%%%%%%%%%%%%%%%%%%%%%%%%%%%%%%%%%%%%%%%%%%%%%%%%%%%%%%%%%%%%%%%%%%%%%%%%%
%%%%%%%%%%%%%%%%%%%%%%%%%%%%%%%%%%%%%%%%%%%%%%%%%%%%%%%%%%%%%%%%%%%%%%%%%%%%%%%%%%%%%%%%%%%%%%%%%%%%%%%%%%%%%%%
\subsection{Spin model~\eqref{eq:model}} \label{ap:w_r}
%%%%%%%%%%%%%%%%%%%%%%%%%%%%%%%%%%%%%%%%%%%%%%%%%%%%%%%%%%%%%%%%%%%%%%%%%%%%%%%%%%%%%%%%%%%%%%%%%%%%%%%%%%%%%%%
%%%%%%%%%%%%%%%%%%%%%%%%%%%%%%%%%%%%%%%%%%%%%%%%%%%%%%%%%%%%%%%%%%%%%%%%%%%%%%%%%%%%%%%%%%%%%%%%%%%%%%%%%%%%%%%

In the case of the spin model~\eqref{eq:model}, we introduce
\begin{equation}
 u_r(t) = \int_0^{ 2 \pi} \frac{dk}{2 \pi} \vr(k,t) e^{-i(k-\bar{k})r}.
\end{equation}
Note that, by construction,
\begin{equation}
    u_r^* = u_{-r} \quad \text{and} \quad w_r = \frac{u_r+u_{-r}}{2} = {\rm Re}[u_r].
\end{equation}
Using  Eq.~\eqref{eq:GGEeqNum}, we can derive an infinite hierarchy of differential equations of the form
\begin{equation}
   \dot{u}_r =  \int_0^{ 2 \pi}\frac{dk}{2 \pi} \frac{d \vr(k)}{dt} e^{-i(k-\bar{k})r} = F_r[\{u_l\}], \label{eq:u_F}
\end{equation}
where $F_r[\{u_l\}]$ is \textit{a priori} a functional of all the $\{ u_l , \ l \in \mathbb{Z}\}$ and which needs to be determined. To do so, we use the expressions of the Hilbert transform and its derivative in Fourier space: 
\begin{equation}
 \mathcal{H}[f](k) = - i \sum_r \text{sgn} (r) f_r e^{i r k}, \qquad \mathcal{H}[f]^\prime(k) =  \sum_r |r| f_r e^{i r k} \label{eq:Hilbert_fourier},
\end{equation}
where
\begin{equation}
    f_r = \int_0^{2 \pi} \frac{dk}{2 \pi} f(k) e^{-i r k}, \qquad f(k) = \sum_{r=-\infty}^\infty f_r e^{irk},
\end{equation}
and ${\rm sgn} (r)=0$, if $r=0$, $1$ if $r>0$ and $-1$, if $r<0$.
After a lengthy but straightforward calculation, we obtain the following system of differential equations:
\begin{subequations}
\begin{equation}
 \frac{1}{\kappa}\dot{u}_0 = u_{-1} - 2 u_0 + u_{1}, \label{eq:du_0}
\end{equation}
 \begin{equation}
\begin{split}
\forall r \neq 0, \ \frac{1}{\kappa} \dot{u}_r =  &   \left( u_{r-1} - 2 u_r + u_{r+1} \right) \left( 1 - u_0\right) +  \text{sgn}(r) u_0 \left( u_{r+1} - u_{r-1}\right) \\ &+ \text{sgn}(r) u_r \left( u_{-1} - u_{1} \right)   +  \left(   2|r| - 1 \right) u_r \frac{\dot{u}_0}{\kappa} - \mathcal{R}(r),
\end{split}
\end{equation} \label{eq:du}
\end{subequations}
with
\begin{equation}
\mathcal{R}(r) =
 \begin{cases}
 2 \displaystyle\sum_{l=1}^{r-1} u_{r- l} \left( u_{l-1} - 2 u_{ l } + u_{l+1}\right), \ \text{if} \ r \geq 2,\\
  2 \displaystyle\sum_{l=r+1}^{-1} u_{r- l} \left( u_{l-1} - 2 u_{ l } + u_{l+1}\right), \ \text{if} \ r \leq -2,\\
 0,  \ \text{otherwise}.
 \end{cases}
\end{equation}
Eq.~\eqref{eq:dw0} is a direct consequence of Eq.~\eqref{eq:du_0}, while Eq.~\eqref{eq:dM} is found by summing Eqs.~\eqref{eq:du} over all $r \in \mathbb{Z}$.

\subsubsection{Derivation of Eqs.~\eqref{eq:u0_delta}}
We now derive Eqs.~\eqref{eq:u0_delta}.
By substituting the ansatz~\eqref{eq:Ansatz} into Eq.~\eqref{eq:dw0}, we obtain
\begin{equation}
 \dot{w}_0  = - 2 \kappa w_0 \left[ 1 - f\left(\frac{1}{\ww}\right) \right].
\end{equation}
Importantly, the characteristic width of the gaussian profile $\ww(t)$ becomes really large at late times, therefore
\begin{equation}
 f\left(\frac{1}{\ww}\right) = \underbrace{f(0)}_{=1} + \frac{1}{\ww} \underbrace{f'(0)}_{=0} + \frac{1}{2 \ww^2} f''(0) + \ldots,
\end{equation}
and we have
\begin{equation}
\dot{w}_0  \simeq \kappa f''(0) \frac{w_0}{\ww^2}. \label{eq:du_02}
\end{equation}
In order to find a dynamical equation for $\ww(t)$, we use Eq.~\eqref{eq:dM}.
We have
\begin{equation}
 M = w_0 \sum_{r=-\infty}^{\infty} f \left( \frac{r}{\ww}\right), \quad \text{and} \quad K = w_0 \sum_{r=-\infty}^{\infty} |r| f \left( \frac{r}{\ww}\right).
\end{equation}
Since $\ww \gg 1$, we can approximate these sums with integrals.
On the one hand, we obtain
\begin{equation}
 \sum_{r=-\infty}^\infty f\left( \frac{r}{\ww} \right) \simeq \int_{-\infty}^\infty f\left( \frac{r}{\ww} \right) dr = \ww \underbrace{\int_{-\infty}^\infty f(x) dx}_{=I} ,
\end{equation}
so that,
\begin{equation}
 M \simeq I w_0 \ww. \label{eq:M2}
\end{equation}
On the other hand, we find
\begin{equation}
 \sum_{r=-\infty}^\infty |r| f\left( \frac{r}{\ww} \right) \simeq \int_{-\infty}^\infty |r| f\left( \frac{r}{\ww} \right) dr = \ww^2 \underbrace{\int_{-\infty}^{\infty} |x| f(x) dx}_{=J},
\end{equation}
meaning that
\begin{equation}
  K \simeq J  w_0 \ww^2 . \label{eq:K}
\end{equation}
Therefore by substituting Eqs.~\eqref{eq:du_02},~\eqref{eq:M2},~\eqref{eq:K} into  Eq.~\eqref{eq:dM}, we find
\begin{equation}
  \frac{1}{\kappa} \dot{\ww} \simeq - \frac{ f''(0) }{\ww } +\frac{2 J f''(0) - 4}{I}   w_0  . \label{eq:ddelta}
\end{equation}
Since $f(x) = \exp(-x^2)$, we have
\begin{equation}
f''(0)=-2, \quad I = \sqrt{\pi}, \quad \text{and}, \quad J = 1. \label{eq:f_I_J}
\end{equation}
Thus, from Eqs.~\eqref{eq:du_02},~\eqref{eq:ddelta} and~\eqref{eq:f_I_J}, we find Eqs.~\eqref{eq:u0_delta}.

\subsubsection{Derivation of Eqs.~\eqref{eq:delta_of_u0}}

We now derive Eq.~\eqref{eq:delta_of_u0}. From Eq.~\eqref{eq:u0_delta}, we have
\begin{equation}
 \frac{d \ww}{d w_0} = \frac{\dot{\ww}}{\dot{w}_0} = - \frac{\ww}{w_0} + \frac{4}{\sqrt{\pi}} \ww^2.
\end{equation}
We introduce the change of variable $y=1/\ww$, thus we have the linear first-order ordinary differential equation 
\begin{equation}
 \frac{d y}{d w_0} =  \frac{y}{w_0} - \frac{4}{\sqrt{\pi}},
\end{equation}
whose solution is 
\begin{equation}
    y = A w_0 - \frac{4}{\sqrt{\pi}} w_0 \log(w_0),
\end{equation}
where $A$ is a constant. Therefore, we find Eq.~\eqref{eq:delta_of_u0}.

%%%%%%%%%%%%%%%%%%%%%%%%%%%%%%%%%%%%%%%%%%%%%%%%%%%%%%%%%%%%%%%%%%%%%%%%%%%%%%%%%%%%%%%%%%%%%%%%%%%%%%%%%%%%%%%
%%%%%%%%%%%%%%%%%%%%%%%%%%%%%%%%%%%%%%%%%%%%%%%%%%%%%%%%%%%%%%%%%%%%%%%%%%%%%%%%%%%%%%%%%%%%%%%%%%%%%%%%%%%%%%%
\section{Details on Eqs.~\eqref{eq:dn_ana} and~\eqref{eq:n_asympotic}} \label{ap:Link_MM}
%%%%%%%%%%%%%%%%%%%%%%%%%%%%%%%%%%%%%%%%%%%%%%%%%%%%%%%%%%%%%%%%%%%%%%%%%%%%%%%%%%%%%%%%%%%%%%%%%%%%%%%%%%%%%%%
%%%%%%%%%%%%%%%%%%%%%%%%%%%%%%%%%%%%%%%%%%%%%%%%%%%%%%%%%%%%%%%%%%%%%%%%%%%%%%%%%%%%%%%%%%%%%%%%%%%%%%%%%%%%%%%

In this appendix, we show that the infinite-time asymptotic solution of Eq.~\eqref{eq:dn_ana} is indeed the result found in Ref.~\cite{daraban2026} recalled in Eq.~\eqref{eq:n_asympotic}.

From Eq.~\eqref{eq:dn_ana}, we find
\begin{equation}
    - \int_{n{(t_0)}}^n\frac{ds}{s^3 \left( C - 4 \sqrt{\frac{2}{\pi}} \log(s)\right)^2} =  \kappa (t-t_0),
\end{equation}
which implies
\begin{equation}
 \frac{\pi}{16} e^{-\sqrt{\frac{\pi}{8}} C } {\rm Ei}\left( \sqrt{\frac{\pi}{8}} C  - 2 \log(n) \right) - \frac{\pi}{16 n^2}  \frac{1}{\sqrt{\frac{\pi}{8}} C  - 2 \log(n)}  + F  =  \kappa t.
\end{equation}
Here, ${\rm Ei}(x) = \fint_{-\infty}^x \frac{e^s}{s} ds$ is the exponential integral, and $F$ is a constant of integration.
However, at infinite times, we have $n \rightarrow0^+$, therefore  $\sqrt{\frac{\pi}{8}} C  - 2 \log(n) \rightarrow \infty$. We can thus use the asymptotic expansion
\begin{equation}
    {\rm Ei}(x) \underset{x \rightarrow  \infty}{\simeq} \frac{e^x}{x} \left( 1+ \frac{1}{x} + \ldots \right)
\end{equation}
to obtain
\begin{equation}
  \frac{\pi}{16}n^{-2} \left(  \sqrt{\frac{\pi}{8}} C - 2 \log(n) \right)^{-2}  +F  \simeq  \kappa t.\label{eq:relation_n_t}
\end{equation}
The relation~\eqref{eq:relation_n_t} can be inverted as
\begin{equation}
 n(t) \simeq - \frac{\sqrt{\pi}/8}{ W_{-1} \left( \frac{G}{\sqrt{\kappa t - F}} \right) \sqrt{\kappa t - F}}, \label{eq:n_W}
\end{equation}
where $G = - \frac{\sqrt{\pi}}{8} \exp(-\frac{\sqrt{\pi} C}{4 \sqrt{2}})<0$. Here, $W_{-1}$ is the Lambert $W$ function for the $-1$ branch; it satisfies by definition $W_{-1}(z)  \exp\left(W_{-1}(z)\right)=z$.

In addition, when $t \rightarrow \infty$, we can use the asymptotic expansion
\begin{equation}
 W_{-1}(x) \underset{x \rightarrow 0^{-}}{\simeq} \log(-x) - \log \left(- \log(-x)\right)  + \ldots \label{eq:exp_W}
\end{equation}
Therefore, at leading order
\begin{equation}
    W_{-1} \left( \frac{G}{\sqrt{\kappa t - F}} \right) \sqrt{\kappa t - F} \underset{t \rightarrow \infty} \sim -\frac{1}{2} \log(\kappa t) \sqrt{\kappa t}. \label{eq:approx_W}
\end{equation}
Eqs.~\eqref{eq:n_W} and~\eqref{eq:approx_W} lead to Eq.~\eqref{eq:n_asympotic}. Note that according to the expansion of the Lambert $W$ function~\eqref{eq:exp_W}, a higher order term of the form $\log(\log(t))$ could be added in the denominator of Eq.~\eqref{eq:n_asympotic}.

\end{appendix}

\bibliography{biblio.bib}

@Article{Avni2025,
  title = {{Dynamical phase transitions in the nonreciprocal Ising model}},
  author = {Avni, Yael and Fruchart, Michel and Martin, David and Seara, Daniel and Vitelli, Vincenzo},
  journal = {Phys. Rev. E},
  volume = {111},
  issue = {3},
  pages = {034124},
  numpages = {41},
  year = {2025},
  month = {Mar},
  publisher = {American Physical Society},
  doi = {10.1103/PhysRevE.111.034124},
  url = {https://link.aps.org/doi/10.1103/PhysRevE.111.034124}
}

@article{Metelmann2015,
  title = {{Nonreciprocal Photon Transmission and Amplification via Reservoir Engineering}},
  author = {Metelmann, A. and Clerk, A. A.},
  journal = {Phys. Rev. X},
  volume = {5},
  issue = {2},
  pages = {021025},
  numpages = {16},
  year = {2015},
  month = {Jun},
  publisher = {American Physical Society},
  doi = {10.1103/PhysRevX.5.021025},
  url = {https://link.aps.org/doi/10.1103/PhysRevX.5.021025}
}

@article{Metelmann2017,
  title = {Nonreciprocal quantum interactions and devices via autonomous feedforward},
  author = {Metelmann, A. and Clerk, A. A.},
  journal = {Phys. Rev. A},
  volume = {95},
  issue = {1},
  pages = {013837},
  numpages = {9},
  year = {2017},
  month = {Jan},
  publisher = {American Physical Society},
  doi = {10.1103/PhysRevA.95.013837},
  url = {https://link.aps.org/doi/10.1103/PhysRevA.95.013837}
}

@Article{Clerk2022,
	title={{Introduction to quantum non-reciprocal interactions: from non-Hermitian Hamiltonians to quantum master equations and quantum feedforward schemes}},
	author={Aashish A. Clerk},
	journal={SciPost Phys. Lect. Notes},
	pages={44},
	year={2022},
	publisher={SciPost},
	doi={10.21468/SciPostPhysLectNotes.44},
	url={https://scipost.org/10.21468/SciPostPhysLectNotes.44},
}

@misc{Soares2025,
      title={Dissipative phase transition of interacting non-reciprocal fermions}, 
      author={Rafael D. Soares and Matteo Brunelli and Marco Schirò},
      year={2025},
      eprint={2505.15711},
      archivePrefix={arXiv},
      primaryClass={quant-ph},
      url={https://arxiv.org/abs/2505.15711}, 
}

@Article{Fazio2025,
	title={{Many-body open quantum systems}},
	author={Rosario Fazio and Jonathan Keeling and Leonardo Mazza and Marco Schirò},
	journal={SciPost Phys. Lect. Notes},
	pages={99},
	year={2025},
	publisher={SciPost},
	doi={10.21468/SciPostPhysLectNotes.99},
	url={https://scipost.org/10.21468/SciPostPhysLectNotes.99},
}

@article{Ashida_2020,
   title={{Non-Hermitian physics}},
   volume={69},
   ISSN={1460-6976},
   url={http://dx.doi.org/10.1080/00018732.2021.1876991},
   DOI={10.1080/00018732.2021.1876991},
   number={3},
   journal={Advances in Physics},
   publisher={Informa UK Limited},
   author={Ashida, Yuto and Gong, Zongping and Ueda, Masahito},
   year={2020},
   month=jul, pages={249–435} }

@article{Song_2019,
  title = {{Non-Hermitian Skin Effect and Chiral Damping in Open Quantum Systems}},
  author = {Song, Fei and Yao, Shunyu and Wang, Zhong},
  journal = {Phys. Rev. Lett.},
  volume = {123},
  issue = {17},
  pages = {170401},
  numpages = {8},
  year = {2019},
  month = {Oct},
  publisher = {American Physical Society},
  doi = {10.1103/PhysRevLett.123.170401},
  url = {https://link.aps.org/doi/10.1103/PhysRevLett.123.170401}
}

@Article{Brunelli_2023,
	title={{Restoration of the non-Hermitian bulk-boundary correspondence via topological amplification}},
	author={Matteo Brunelli and Clara C. Wanjura and Andreas Nunnenkamp},
	journal={SciPost Phys.},
	volume={15},
	pages={173},
	year={2023},
	publisher={SciPost},
	doi={10.21468/SciPostPhys.15.4.173},
	url={https://scipost.org/10.21468/SciPostPhys.15.4.173},
}

@article{Porras_2019,
  title = {{Topological Amplification in Photonic Lattices}},
  author = {Porras, Diego and Fern\'andez-Lorenzo, Samuel},
  journal = {Phys. Rev. Lett.},
  volume = {122},
  issue = {14},
  pages = {143901},
  numpages = {6},
  year = {2019},
  month = {Apr},
  publisher = {American Physical Society},
  doi = {10.1103/PhysRevLett.122.143901},
  url = {https://link.aps.org/doi/10.1103/PhysRevLett.122.143901}
}

@article{Wanjura_2020,
  title = {Topological framework for directional amplification in driven-dissipative cavity arrays},
  author = {Wanjura, Clara C. and Brunelli, Matteo and Nunnenkamp, Andreas},
  journal = {Nature Communications},
  volume = {11},
  pages = {3149},
  year = {2020},
  doi = {10.1038/s41467-020-16863-9},
}

@article{Essler_2016,
   title={Quench dynamics and relaxation in isolated integrable quantum spin chains},
   volume={2016},
   ISSN={1742-5468},
   url={http://dx.doi.org/10.1088/1742-5468/2016/06/064002},
   DOI={10.1088/1742-5468/2016/06/064002},
   number={6},
   journal={Journal of Statistical Mechanics: Theory and Experiment},
   publisher={IOP Publishing},
   author={Essler, Fabian H L and Fagotti, Maurizio},
   year={2016},
   month=jun, pages={064002} }

@article{VidmarRigol2016,
  author       = {Lev Vidmar and Marcos Rigol},
  title        = {{Generalized Gibbs ensemble in integrable lattice models}},
  journal      = {Journal of Statistical Mechanics: Theory and Experiment},
  year         = {2016},
  issue        = {06},
  pages        = {064007},
  doi          = {10.1088/1742-5468/2016/06/064007},
}

@article{Reiter_2021,
  title = {{Engineering generalized Gibbs ensembles with trapped ions}},
  author = {Reiter, Florentin and Lange, Florian and Jain, Shreyans and Grau, Matt and Home, Jonathan P. and Lenar\ifmmode \check{c}\else \v{c}\fi{}i\ifmmode \check{c}\else \v{c}\fi{}, Zala},
  journal = {Phys. Rev. Res.},
  volume = {3},
  issue = {3},
  pages = {033142},
  numpages = {18},
  year = {2021},
  month = {Aug},
  publisher = {American Physical Society},
  doi = {10.1103/PhysRevResearch.3.033142},
  url = {https://link.aps.org/doi/10.1103/PhysRevResearch.3.033142}
}

@article{Poletti_2012,
  title = {{Interaction-Induced Impeding of Decoherence and Anomalous Diffusion}},
  author = {Poletti, Dario and Bernier, Jean-S\'ebastien and Georges, Antoine and Kollath, Corinna},
  journal = {Phys. Rev. Lett.},
  volume = {109},
  issue = {4},
  pages = {045302},
  numpages = {5},
  year = {2012},
  month = {Jul},
  publisher = {American Physical Society},
  doi = {10.1103/PhysRevLett.109.045302},
  url = {https://link.aps.org/doi/10.1103/PhysRevLett.109.045302}
}

@article{Poletti_2013,
   title={{Emergence of Glasslike Dynamics for Dissipative and Strongly Interacting Bosons}},
   volume={111},
   ISSN={1079-7114},
   url={http://dx.doi.org/10.1103/PhysRevLett.111.195301},
   DOI={10.1103/physrevlett.111.195301},
   number={19},
   journal={Physical Review Letters},
   publisher={American Physical Society (APS)},
   author={Poletti, Dario and Barmettler, Peter and Georges, Antoine and Kollath, Corinna},
   year={2013},
   month=nov }

@article{Perfetto_2023,
   title={{Reaction-Limited Quantum Reaction-Diffusion Dynamics}},
   volume={130},
   ISSN={1079-7114},
   url={http://dx.doi.org/10.1103/PhysRevLett.130.210402},
   DOI={10.1103/physrevlett.130.210402},
   number={21},
   journal={Physical Review Letters},
   publisher={American Physical Society (APS)},
   author={Perfetto, Gabriele and Carollo, Federico and Garrahan, Juan P. and Lesanovsky, Igor},
   year={2023},
   month=may }

@article{Pocklington_2025,
  title = {{Efficient Simulation of Nontrivial Dissipative Spin Chains via Stochastic Unraveling}},
  author = {Pocklington, Andrew and Clerk, Aashish A.},
  journal = {PRX Quantum},
  volume = {6},
  issue = {3},
  pages = {030349},
  numpages = {29},
  year = {2025},
  month = {Sep},
  publisher = {American Physical Society},
  doi = {10.1103/vptq-xy6h},
  url = {https://link.aps.org/doi/10.1103/vptq-xy6h}
}

@Article{Nagy2010,
author={Nagy, M{\'a}t{\'e} and {\'A}kos, Zsuzsa and Biro, Dora and Vicsek, Tam{\'a}s},
title={Hierarchical group dynamics in pigeon flocks},
journal={Nature},
year={2010},
month={Apr},
day={01},
volume={464},
number={7290},
pages={890-893},
issn={1476-4687},
doi={10.1038/nature08891},
url={https://doi.org/10.1038/nature08891}
}

@Article{Tan2022,
author={Tan, Tzer Han and Mietke, Alexander and Li, Junang and Chen, Yuchao and Higinbotham, Hugh and Foster, Peter J. and Gokhale, Shreyas and Dunkel, J{\"o}rn and Fakhri, Nikta},
title={Odd dynamics of living chiral crystals},
journal={Nature},
year={2022},
month={Jul},
day={01},
volume={607},
number={7918},
pages={287-293},
issn={1476-4687},
doi={10.1038/s41586-022-04889-6},
url={https://doi.org/10.1038/s41586-022-04889-6}
}

@Article{Dinelli2023,
author={Dinelli, Alberto and O'Byrne, J{\'e}r{\'e}my and Curatolo, Agnese and Zhao, Yongfeng and Sollich, Peter and Tailleur, Julien},
title={Non-reciprocity across scales in active mixtures},
journal={Nature Communications},
year={2023},
month={Nov},
day={03},
volume={14},
number={1},
pages={7035},
issn={2041-1723},
doi={10.1038/s41467-023-42713-5},
url={https://doi.org/10.1038/s41467-023-42713-5}
}

@article{Vicsek_2012,
   title={Collective motion},
   volume={517},
   ISSN={0370-1573},
   url={http://dx.doi.org/10.1016/j.physrep.2012.03.004},
   DOI={10.1016/j.physrep.2012.03.004},
   number={3–4},
   journal={Physics Reports},
   publisher={Elsevier BV},
   author={Vicsek, Tamás and Zafeiris, Anna},
   year={2012},
   month=aug, pages={71–140} }

@article{johnsrud_2025,
  title = {Fluctuation dissipation relations for active field theories},
  author = {Johnsrud, Martin Kj\o{}llesdal and Golestanian, Ramin},
  journal = {Phys. Rev. Res.},
  volume = {7},
  issue = {3},
  pages = {L032053},
  numpages = {7},
  year = {2025},
  month = {Sep},
  publisher = {American Physical Society},
  doi = {10.1103/xx4z-lj5c},
  url = {https://link.aps.org/doi/10.1103/xx4z-lj5c}
}

@article{Fruchart_2023,
   title={{Odd Viscosity and Odd Elasticity}},
   volume={14},
   ISSN={1947-5462},
   url={http://dx.doi.org/10.1146/annurev-conmatphys-040821-125506},
   DOI={10.1146/annurev-conmatphys-040821-125506},
   number={1},
   journal={Annual Review of Condensed Matter Physics},
   publisher={Annual Reviews},
   author={Fruchart, Michel and Scheibner, Colin and Vitelli, Vincenzo},
   year={2023},
   month=mar, pages={471–510} }

@Article{Fruchart2021,
author={Fruchart, Michel and Hanai, Ryo and Littlewood, Peter B. and Vitelli, Vincenzo},
title={Non-reciprocal phase transitions},
journal={Nature},
year={2021},
month={Apr},
day={01},
volume={592},
number={7854},
pages={363-369},
issn={1476-4687},
doi={10.1038/s41586-021-03375-9},
url={https://doi.org/10.1038/s41586-021-03375-9}
}

@Article{Uchida2010,
  title = {{Synchronization and Collective Dynamics in a Carpet of Microfluidic Rotors}},
  author = {Uchida, Nariya and Golestanian, Ramin},
  journal = {Phys. Rev. Lett.},
  volume = {104},
  issue = {17},
  pages = {178103},
  numpages = {4},
  year = {2010},
  month = {Apr},
  publisher = {American Physical Society},
  doi = {10.1103/PhysRevLett.104.178103},
  url = {https://link.aps.org/doi/10.1103/PhysRevLett.104.178103}
}

@Article{Brandenbourger2019,
author={Brandenbourger, Martin and Locsin, Xander and Lerner, Edan and Coulais, Corentin},
title={Non-reciprocal robotic metamaterials},
journal={Nature Communications},
year={2019},
month={Oct},
day={10},
volume={10},
number={1},
pages={4608},
issn={2041-1723},
doi={10.1038/s41467-019-12599-3},
url={https://doi.org/10.1038/s41467-019-12599-3}
}

@article{Hatano1996,
  title = {{Localization Transitions in Non-Hermitian Quantum Mechanics}},
  author = {Hatano, Naomichi and Nelson, David R.},
  journal = {Phys. Rev. Lett.},
  volume = {77},
  issue = {3},
  pages = {570--573},
  numpages = {0},
  year = {1996},
  month = {Jul},
  publisher = {American Physical Society},
  doi = {10.1103/PhysRevLett.77.570},
  url = {https://link.aps.org/doi/10.1103/PhysRevLett.77.570}
}

@article{Hatano1997,
  title = {{Vortex pinning and non-Hermitian quantum mechanics}},
  author = {Hatano, Naomichi and Nelson, David R.},
  journal = {Phys. Rev. B},
  volume = {56},
  issue = {14},
  pages = {8651--8673},
  numpages = {0},
  year = {1997},
  month = {Oct},
  publisher = {American Physical Society},
  doi = {10.1103/PhysRevB.56.8651},
  url = {https://link.aps.org/doi/10.1103/PhysRevB.56.8651}
}

@article{Gong2018,
  title = {Topological Phases of Non-Hermitian Systems},
  author = {Gong, Zongping and Ashida, Yuto and Kawabata, Kohei and Takasan, Kazuaki and Higashikawa, Sho and Ueda, Masahito},
  journal = {Phys. Rev. X},
  volume = {8},
  issue = {3},
  pages = {031079},
  numpages = {33},
  year = {2018},
  month = {Sep},
  publisher = {American Physical Society},
  doi = {10.1103/PhysRevX.8.031079},
  url = {https://link.aps.org/doi/10.1103/PhysRevX.8.031079}
}

@article{Okuma2020,
  title = {{Topological Origin of Non-Hermitian Skin Effects}},
  author = {Okuma, Nobuyuki and Kawabata, Kohei and Shiozaki, Ken and Sato, Masatoshi},
  journal = {Phys. Rev. Lett.},
  volume = {124},
  issue = {8},
  pages = {086801},
  numpages = {7},
  year = {2020},
  month = {Feb},
  publisher = {American Physical Society},
  doi = {10.1103/PhysRevLett.124.086801},
  url = {https://link.aps.org/doi/10.1103/PhysRevLett.124.086801}
}

@article{Keck2018,
  title = {Persistent currents by reservoir engineering},
  author = {Keck, Maximilian and Rossini, Davide and Fazio, Rosario},
  journal = {Phys. Rev. A},
  volume = {98},
  issue = {5},
  pages = {053812},
  numpages = {10},
  year = {2018},
  month = {Nov},
  publisher = {American Physical Society},
  doi = {10.1103/PhysRevA.98.053812},
  url = {https://link.aps.org/doi/10.1103/PhysRevA.98.053812}
}

@article{Belyansky2025,
  title = {{Phase Transitions in Nonreciprocal Driven-Dissipative Condensates}},
  author = {Belyansky, Ron and Weis, Cheyne and Hanai, Ryo and Littlewood, Peter B. and Clerk, Aashish A.},
  journal = {Phys. Rev. Lett.},
  volume = {135},
  issue = {12},
  pages = {123401},
  numpages = {9},
  year = {2025},
  month = {Sep},
  publisher = {American Physical Society},
  doi = {10.1103/gphr-d1bc},
  url = {https://link.aps.org/doi/10.1103/gphr-d1bc}
}

@article{McDonald2022,
  title = {{Nonequilibrium stationary states of quantum non-Hermitian lattice models}},
  author = {McDonald, A. and Hanai, R. and Clerk, A. A.},
  journal = {Phys. Rev. B},
  volume = {105},
  issue = {6},
  pages = {064302},
  numpages = {19},
  year = {2022},
  month = {Feb},
  publisher = {American Physical Society},
  doi = {10.1103/PhysRevB.105.064302},
  url = {https://link.aps.org/doi/10.1103/PhysRevB.105.064302}
}

@article{Brighi2024,
  title = {{Nonreciprocal dynamics and the non-Hermitian skin effect of repulsively bound pairs}},
  author = {Brighi, Pietro and Nunnenkamp, Andreas},
  journal = {Phys. Rev. A},
  volume = {110},
  issue = {2},
  pages = {L020201},
  numpages = {6},
  year = {2024},
  month = {Aug},
  publisher = {American Physical Society},
  doi = {10.1103/PhysRevA.110.L020201},
  url = {https://link.aps.org/doi/10.1103/PhysRevA.110.L020201}
}

@article{Begg2024,
  title = {{Quantum Criticality in Open Quantum Spin Chains with Nonreciprocity}},
  author = {Begg, Samuel E. and Hanai, Ryo},
  journal = {Phys. Rev. Lett.},
  volume = {132},
  issue = {12},
  pages = {120401},
  numpages = {7},
  year = {2024},
  month = {Mar},
  publisher = {American Physical Society},
  doi = {10.1103/PhysRevLett.132.120401},
  url = {https://link.aps.org/doi/10.1103/PhysRevLett.132.120401}
}

@article{Lange2018,
   title={{Time-dependent generalized Gibbs ensembles in open quantum systems}},
   volume={97},
   ISSN={2469-9969},
   url={http://dx.doi.org/10.1103/PhysRevB.97.165138},
   DOI={10.1103/physrevb.97.165138},
   number={16},
   journal={Physical Review B},
   publisher={American Physical Society (APS)},
   author={Lange, Florian and Lenarčič, Zala and Rosch, Achim},
   year={2018},
   month=apr }

@Article{Bouchoule2020,
	title={{The effect of atom losses on the distribution of rapidities in the one-dimensional Bose gas}},
	author={Isabelle Bouchoule and Benjamin Doyon and Jerome Dubail},
	journal={SciPost Phys.},
	volume={9},
	pages={044},
	year={2020},
	publisher={SciPost},
	doi={10.21468/SciPostPhys.9.4.044},
	url={https://scipost.org/10.21468/SciPostPhys.9.4.044},
}

@article{Riggio2024,
   title={Effects of atom losses on a one-dimensional lattice gas of hard-core bosons},
   volume={109},
   ISSN={2469-9934},
   url={http://dx.doi.org/10.1103/PhysRevA.109.023311},
   DOI={10.1103/physreva.109.023311},
   number={2},
   journal={Physical Review A},
   publisher={American Physical Society (APS)},
   author={Riggio, François and Rosso, Lorenzo and Karevski, Dragi and Dubail, Jérôme},
   year={2024},
   month=feb }

@article{Lehr2025,
doi = {10.1088/1367-2630/adef70},
url = {https://doi.org/10.1088/1367-2630/adef70},
year = {2025},
month = {aug},
publisher = {IOP Publishing},
volume = {27},
number = {8},
pages = {084602},
author = {Lehr, Hannah and Lesanovsky, Igor and Perfetto, Gabriele},
title = {{Reaction-diffusion dynamics of the weakly dissipative Fermi gas}},
journal = {New Journal of Physics}
}

@Article{Ulcakar2025,
	title={{Generalized Gibbs ensembles in weakly interacting dissipative systems and digital quantum computers}},
	author={Iris Ulčakar and Zala Lenarčič},
	journal={SciPost Phys.},
	volume={19},
	pages={068},
	year={2025},
	publisher={SciPost},
	doi={10.21468/SciPostPhys.19.3.068},
	url={https://scipost.org/10.21468/SciPostPhys.19.3.068},
}

@article{Lumia2025,
  title = {{Accuracy of a time-dependent generalized Gibbs ensemble approach under weak dissipation}},
  author = {Lumia, Luca and Aupetit-Diallo, Gianni and Dubail, J\'er\^ome and Collura, Mario},
  journal = {Phys. Rev. A},
  volume = {112},
  issue = {1},
  pages = {012206},
  numpages = {16},
  year = {2025},
  month = {Jul},
  publisher = {American Physical Society},
  doi = {10.1103/x9c1-hyxh},
  url = {https://link.aps.org/doi/10.1103/x9c1-hyxh}
}

@article{Cai2013,
  title = {{Algebraic versus Exponential Decoherence in Dissipative Many-Particle Systems}},
  author = {Cai, Zi and Barthel, Thomas},
  journal = {Phys. Rev. Lett.},
  volume = {111},
  issue = {15},
  pages = {150403},
  numpages = {5},
  year = {2013},
  month = {Oct},
  publisher = {American Physical Society},
  doi = {10.1103/PhysRevLett.111.150403},
  url = {https://link.aps.org/doi/10.1103/PhysRevLett.111.150403}
}

@article{Rossini2021,
  title = {{Strong correlations in lossy one-dimensional quantum gases: From the quantum Zeno effect to the generalized Gibbs ensemble}},
  author = {Rossini, Davide and Ghermaoui, Alexis and Aguilera, Manel Bosch and Vatr\'e, R\'emy and Bouganne, Rapha\"el and Beugnon, J\'er\^ome and Gerbier, Fabrice and Mazza, Leonardo},
  journal = {Phys. Rev. A},
  volume = {103},
  issue = {6},
  pages = {L060201},
  numpages = {5},
  year = {2021},
  month = {Jun},
  publisher = {American Physical Society},
  doi = {10.1103/PhysRevA.103.L060201},
  url = {https://link.aps.org/doi/10.1103/PhysRevA.103.L060201}
}

@article{Marche2024,
  title = {{Universality and two-body losses: Lessons from the effective non-Hermitian dynamics of two particles}},
  author = {March\'e, Alice and Yoshida, Hironobu and Nardin, Alberto and Katsura, Hosho and Mazza, Leonardo},
  journal = {Phys. Rev. A},
  volume = {110},
  issue = {3},
  pages = {033321},
  numpages = {20},
  year = {2024},
  month = {Sep},
  publisher = {American Physical Society},
  doi = {10.1103/PhysRevA.110.033321},
  url = {https://link.aps.org/doi/10.1103/PhysRevA.110.033321}
}

@Article{Rosso2022,
	title={{The one-dimensional Bose gas with strong two-body losses: the effect of the harmonic confinement}},
	author={Lorenzo Rosso and Alberto Biella and Leonardo Mazza},
	journal={SciPost Phys.},
	volume={12},
	pages={044},
	year={2022},
	publisher={SciPost},
	doi={10.21468/SciPostPhys.12.1.044},
	url={https://scipost.org/10.21468/SciPostPhys.12.1.044},
}

@inbook{King2009, place={Cambridge}, series={Encyclopedia of Mathematics and its Applications}, title={The Hilbert transform of periodic functions}, booktitle={Hilbert Transforms}, publisher={Cambridge University Press}, author={King, Frederick W.}, year={2009}, pages={288–330}, collection={Encyclopedia of Mathematics and its Applications}}

@article{Pandey1997,
author = {J. N. Pandey},
title = {{The Hilbert transform of periodic distributions}},
journal = {Integral Transforms and Special Functions},
volume = {5},
number = {1-2},
pages = {117--142},
year = {1997},
publisher = {Taylor \& Francis},
doi = {10.1080/10652469708819129},
URL = {https://doi.org/10.1080/10652469708819129},
eprint = {https://doi.org/10.1080/10652469708819129}}

@article{Johansson2012,
title = {{QuTiP: An open-source Python framework for the dynamics of open quantum systems}},
journal = {Computer Physics Communications},
volume = {183},
number = {8},
pages = {1760-1772},
year = {2012},
issn = {0010-4655},
doi = {https://doi.org/10.1016/j.cpc.2012.02.021},
url = {https://www.sciencedirect.com/science/article/pii/S0010465512000835},
author = {J.R. Johansson and P.D. Nation and Franco Nori}
}

@article{Johansson2013,
title = {{QuTiP 2: A Python framework for the dynamics of open quantum systems}},
journal = {Computer Physics Communications},
volume = {184},
number = {4},
pages = {1234-1240},
year = {2013},
issn = {0010-4655},
doi = {https://doi.org/10.1016/j.cpc.2012.11.019},
url = {https://www.sciencedirect.com/science/article/pii/S0010465512003955},
author = {J.R. Johansson and P.D. Nation and Franco Nori}
}

@misc{daraban2026,
      title={Universal dynamics from a single-particle dark state}, 
      author={Ruben Daraban and Arghavan Safavi-Naini and Johannes Schachenmayer and Mohammad Maghrebi},
      year={2026},
      eprint={2605.16494},
      archivePrefix={arXiv},
      primaryClass={cond-mat.quant-gas},
      url={https://arxiv.org/abs/2605.16494}, 
}

@article{Yamamoto2020,
  title = {Rectification in nonequilibrium steady states of open many-body systems},
  author = {Yamamoto, Kazuki and Ashida, Yuto and Kawakami, Norio},
  journal = {Phys. Rev. Res.},
  volume = {2},
  issue = {4},
  pages = {043343},
  numpages = {12},
  year = {2020},
  month = {Dec},
  publisher = {American Physical Society},
  doi = {10.1103/PhysRevResearch.2.043343},
  url = {https://link.aps.org/doi/10.1103/PhysRevResearch.2.043343}
}

@article{Perfetto_2023_2,
   title={Quantum reaction-limited reaction-diffusion dynamics of annihilation processes},
   volume={108},
   ISSN={2470-0053},
   url={http://dx.doi.org/10.1103/PhysRevE.108.064104},
   DOI={10.1103/physreve.108.064104},
   number={6},
   journal={Physical Review E},
   publisher={American Physical Society (APS)},
   author={Perfetto, Gabriele and Carollo, Federico and Garrahan, Juan P. and Lesanovsky, Igor},
   year={2023},
   month=Dec }

@article{Medvedyeva_2014,
  title = {Power-law approach to steady state in open lattices of noninteracting electrons},
  author = {Medvedyeva, M. V. and Kehrein, S.},
  journal = {Phys. Rev. B},
  volume = {90},
  issue = {20},
  pages = {205410},
  numpages = {18},
  year = {2014},
  month = {Nov},
  publisher = {American Physical Society},
  doi = {10.1103/PhysRevB.90.205410},
  url = {https://link.aps.org/doi/10.1103/PhysRevB.90.205410}
}

@misc{hosseinabadi2026,
      title={Transient and universal regimes in quantum reaction-transport kinetics}, 
      author={Hossein Hosseinabadi and Roderich Moessner},
      year={2026},
      eprint={2609.09305},
      archivePrefix={arXiv},
      primaryClass={cond-mat.stat-mech},
      url={https://arxiv.org/abs/2609.09305}, 
}

\nolinenumbers

\end{document}